\documentclass[12pt,oneside,english]{amsart}
\usepackage[T1]{fontenc}
\usepackage[latin9]{inputenc}
\usepackage{color}
\usepackage{amstext}
\usepackage{amsthm}
\usepackage{amssymb}
\usepackage{setspace}
\usepackage[authoryear,round]{natbib}
\usepackage{graphicx}
\makeatletter
\theoremstyle{plain}
\newtheorem{lem}{\protect\lemmaname}
\theoremstyle{definition}
\newtheorem{defn}{\protect\definitionname}
\theoremstyle{plain}
\newtheorem{thm}{\protect\theoremname}
\theoremstyle{plain}
\newtheorem{cor}{\protect\corollaryname}

\usepackage{lmodern}
\usepackage[T1]{fontenc}
\usepackage{braket}
\usepackage{hyperref}
\hypersetup{
colorlinks=false,
citecolor=green
}

\makeatother

\usepackage{babel}
\providecommand{\corollaryname}{Corollary}
\providecommand{\definitionname}{Definition}
\providecommand{\lemmaname}{Lemma}
\providecommand{\theoremname}{Theorem}

\begin{document}
\title{Equilibrium Information Aggregation under Machine Learning}

\author{Andrew Ellis, Michele Piccione, Shengxing Zhang}

\thanks{Ellis: LSE, a.ellis@lse.ac.uk. Piccione: LSE, m.piccione@lse.ac.uk. Zhang: CMU, shengxiz@andrew.cmu.edu. The authors wish to thank Yoram Halevy, Christian Hellwig, Tai-wei Hu, Fei Li, Matt Levy,  Rani Spiegler, and Jakub Steiner, as well as conference/seminar participants at BRIC, Bristol, CBS, CMU finance, Duke/UNC, Leicester, KCL, Makris Symposium, Peking University, QMUL, Southampton, Manchester,  Toronto, and UCL}

\begin{abstract}
We introduce a framework for studying the equilibrium effects of machine
learning. Agents process information using a \citet{chow1968approximating}
tree, a widely-used machine learning procedure that admits a closed-form
solution. We apply the model to an asset market with dispersed information
based on \cite{hellwig1980aggregation}. The price mechanism fails
to aggregate the information extracted by the algorithm, even approximately.
While there are partial equilibrium benefits from access to algorithms,
the equilibrium price aggregates less information than the rational equilibrium. Equilibrium typically features diverse world-models, demands, and utilities, even with ex ante identical agents.
\end{abstract}

\maketitle

\section{Introduction}
The large  and increasing amount of information available in the economy makes it difficult to behave as a good Bayesian would. Machine learning algorithms process this information and extract useful predictions that humans alone would not. 
In an individual decision, this unambiguously increases the information available to the decision maker. In markets and other interactive decisions, the effects are less clear.
As more market participants rely on algorithms, the distribution of endogenous, informative variables like prices changes to ensure equilibrium. These variables aggregate information across individuals, so the overall informativeness depends on the feedback between how the information extracted by algorithm from them affects their distribution, and how the distribution of the endogenous variables affect the information extracted by the algorithm.  This paper seeks to understand the consequences for how the equilibrium price aggregates information. 

We develop a framework for analyzing competitive equilibria when agents form beliefs from an endogenous price using a machine learning algorithm. Specifically, agents use the \citet{chow1968approximating} Tree  (CLT) algorithm instead of Bayesian updating. CLT  approximates the true distribution with one that factorizes according to a tree-structured Bayesian network, and chooses the approximation that minimizes the Kullback-Leibler divergence from  the truth.  Unlike neural networks or random forests, CLT has a closed-form solution, allowing us to characterize the equilibrium explicitly. 

We apply our framework to a canonical model of learning from a price, the asset market of \cite{hellwig1980aggregation} in which risk-averse traders use both a private signal and the market price to learn about a risky asset's return. The only departure is that traders form beliefs with a CLT calibrated to the equilibrium price distribution rather than by Bayes' rule. This  leads to a competitive equilibrium with markedly different properties.
\cite{hellwig1980aggregation} shows that with Bayesian traders, the equilibrium price is \emph{approximately informationally efficient}:  as uncertainty about supply vanishes, prices reveal all private information. We show that in our framework, even approximate informational efficiency is \emph{impossible}: correlation between price and value is bounded strictly below that in the Bayesian limit.

Section \ref{sec: CLT intro} introduces CLT and justifies our use of it to model machine learning. Bayesian updating is computationally very demanding \citep[formally, NP-hard; see][]{COOPER1990computational}, and CLT simplifies it by approximating a complex joint probability distribution with a simpler  one. This solves the key computational difficulty, the curse of dimensionality. While a full joint distribution over many variables requires tracking an enormous number of parameters (exponential in the number of variables for discrete data), a tree-structured distribution requires only tracking pairwise relationships along the branches of the tree. \citet[][p 321]{jiao2016universal} note that the CLT is still ``widely used in statistics and machine learning as a tool for dimensionality reduction, classification,  and as a foundation for algorithm design in more complex dependence structures.''
%

The CLT algorithm has several features that make it attractive as a model of machine learning. First, it is computationally efficient, running in polynomial time. Second, it produces sparse models: a small subset of relationships suffices for prediction.  Third, the algorithm is non-parametric in the sense that it does not require the analyst to specify a functional form for relationships. Last and crucially for our purposes, the optimal tree has a closed-form characterization, enabling equilibrium analysis. The main limitation relative to richer machine learning methods is that CLTs restrict each variable to have at most one ``parent'' in the dependence structure, precluding the deeper architectures of neural networks. However, finding optimal higher-order trees is NP-hard \citep{chickering1996learning}, which provides a computational justification for its focus  on  trees.

Section \ref{sec:Asset market} embeds CLT in a market setting with dispersed information and an endogenous price. Each trader observes the price and a private signal about the value of a risky asset, then submits a demand for the asset. Traders have CARA utility over terminal wealth, and market clearing requires aggregate demand  to equal a random supply. Each trader's algorithm processes training data consisting of all the  variables that the trader eventually observes: the realized asset value, the market price, and the trader's private signals. Given its equilibrium joint distribution, the algorithm constructs an approximation for it. The trader then updates the approximation using her private information and the public price, and submits the demand that maximizes expected utility.

A CLT equilibrium thus consists of a distribution of prices as well as  (a distribution over)  trees for each trader so that the following three conditions hold. First, each trader's demand must maximize utility given her tree and information. Second, each trader's tree must be a Chow-Liu Tree, that is, the best approximation to the equilibrium joint distribution. Third, the market must clear for every realization of signals, price, and supply shock. 

Section \ref{sec:analysis} analyzes the CLT equilibrium in the asset market. Our first result shows that an equilibrium exists. The main new consideration is the relationship between the tree selected by the algorithm and the market clearing price. The former determines how the trader's demand responds to information. When more traders use trees that link a private signal to the fundamental value, they trade more aggressively on that signal, so the price must be more responsive to it. Ceteris paribus, the price becomes more correlated with it as well as with the underlying value. This changes the quality of the approximation that each tree yields, feeding back into which one the algorithm selects. In equilibrium, the distribution of the price that clears the market for the trees in the population makes each a CLT. We show existence using a nested fixed point argument.

We then turn to the question of how the CLT equilibrium differs from that in \cite{hellwig1980aggregation}. The key finding is that the equilibrium correlation between the price and the value is bounded away from the Bayesian benchmark by an amount that does not vanish even as supply noise goes to zero.  The logic runs as follows. Suppose, toward a contradiction, that the price were sufficiently correlated with the value to reveal (almost) all private information. By the data-processing inequality, the price would also be more correlated with each private signal than that signal is with the value. The best approximating tree then connects all signals to the price rather than to the value. But traders using such a  tree would condition their demand only on price, effectively ignoring their private signals. With no one trading on private information, the market cannot clear, as demand would not vary with the supply shock. 

Equilibrium adjusts the distribution of trees to maintain market clearing. Some traders must use trees that incorporate private signals, which requires that the price not be too informative. This gives an upper bound on price informativeness that does not depend on the level of noise trading, a stark departure from \cite{grossman1976efficiency} and \cite{hellwig1980aggregation}. This  suggests a non-vanishing private value of machine learning: even in liquid markets with minimal noise, traders who deploy sophisticated algorithms retain an informational advantage that is not competed away through prices. Since training such algorithms is costly, this provides an economic rationale for the persistent investment in algorithmic trading that we observe. The result also speaks to the \cite{grossman1980impossibility} paradox, namely  why traders invest in information acquisition when prices should reveal that information. In our framework, the paradox dissolves: traders have a persistent incentive to acquire and process information because the price never fully reveals it. In addition, ex ante identical traders may use different CLTs in equilibrium, leading to asymmetric reactions to the same information. With purely public but high-dimensional information, traders may still update beliefs in response to price changes even though prices are merely a garbling of the public signals.

Section \ref{sec:examples} specializes the model to illustrate some notable features and more easily relate to Bayesian benchmarks. The CLT equilibrium exhibits comparative statics that often reverse the predictions of rational expectations models. In \cite{hellwig1980aggregation}, increasing the number of traders improves price informativeness through diversification of private information; in our model, increasing the number of traders can \emph{reduce} welfare as traders increasingly rely on the (crowded) price signal rather than diverse private information. Similarly, while \cite{grossman1980impossibility} predict that higher risk aversion reduces price informativeness, in our framework higher risk aversion keeps informativeness constant, and simply shifts traders toward trees that incorporate private signals. 

Two strands of papers study economic effects of machine learning or artificial intelligence. First, \cite{calvano2020artificial,klein2021autonomous,banchio2022artificial,dolgopolov2024reinforcement} and others show that reinforcement learning, a different form of machine learning, leads to non-Nash outcomes in pricing games and the prisoner's dilemma.%
\footnote{\cite{banchio2022artificial} argue this is due ``spontaneous coupling,'' an endogenous statistical linkage between variables.} Methodologically, these papers are essentially experimental with the algorithm taking the role of subjects.  
Second, \cite{ely2023natural,liang2026clones,fudenberg2026friend,bergemann2026menu,spiegler2026machine}   study economic models of algorithms with stylized capabilities.  In contrast to these papers, we analytically characterize  the  equilibrium outcome where the updating is performed by an actual algorithm.  The only other paper of which we are aware that analytically solves for equilibrium with an existing algorithm  is \cite{jehiel2026endogenous}, which uses the K-means clustering algorithm to define analogy classes. 

\cite{eliaz2021cheating} is the only economics paper of which we are aware that explicitly uses CLT. They show that the Chow-Liu tree will select the tree that gets maximally tricked.
Since the algorithm potentially introduces misspecification, our paper is also related to the recent literature on misspecified models \citep[e.g.,][]{spiegler2016bayesian,espondapouzo2016,fudenbergEtAl2017,heidhuesEtAl2021,bohrenHauser2021,frickEtAl2022}. The CLT selects the best misspecification. This makes it particularly related to papers like \cite{ChoKasa2015} and \cite{Ba2024} that incorporate model selection and misspecification tests. 

\section{Chow-Liu Trees}
\label{sec: CLT intro}
A directed acyclic graph (DAG) $G=(N,R)$ is a set of nodes $N$ and
edges $R\subset N\times N$ with no directed cycles. We write $xRy$
for $(x,y)\in R$. Consider an $n$-dimensional random variable $X=\left(X_{1},\dots,X_{n}\right)$.
A $K$-order dependence tree is a DAG $\left(\left\{ 1,\dots,n\right\} ,R\right)$,
sometimes just denoted $R$, with the following two properties. First,
if $iRj$ and $kRj$ then either $iRk$ or $kRi$. Second, $|R(i)|\leq K$
for every node $i$ where $R(i)=\left\{ k:kRi\right\} $. The dependence
tree $R$ determines an approximation $\nu_{R}$ of a distribution
$\nu$ using the formula 
\[
\nu_{R}(x_{1},\dots x_{n})=\prod_{j=1}^{n}\nu\left(x_{j}|x_{R(j)}\right)
\]
where $x_{\emptyset}$ is an arbitrary constant, and $\nu_{R},\nu$
are the probability density (or mass) functions of $\nu_{R}$ and
$\nu$, respectively. For example, the tree $R=1\rightarrow 2 \rightarrow 3 \rightarrow 4$ approximates the distribution $\nu$ according to \[\nu _R\left(x_1,x_2,x_3,x_4 \right)=\nu \left(x_1 \right)\nu \left(x_2|x_1 \right)\nu \left(x_3|x_2 \right)\nu \left(x_4|x_3 \right)\]
and thus imposes that $X_3$ is independent of $X_1$ given $X_2$, regardless of their true relationship.

Dependence trees have several computational advantages. First, the
set 
\[
\left\{ \nu_{R}:\nu\in\Delta\left(\mathcal{X}_{1}\times\dots\times\mathcal{X}_{n}\right)\right\} 
\]
has lower dimension than $\Delta\left(\mathcal{X}_{1}\times\dots\times\mathcal{X}_{n}\right)$.
Second $\nu_{R}$ is fully determined by the marginal distribution
on the cliques of $R$ (subsets of nodes on which $R$ is complete).
Third, it leads to simple and computationally quicker updating.
These capture salient and important features of modern machine learning
models. 

In our analysis, we use the \cite{chow1968approximating} Tree algorithm (CLT) to select a 1-order dependence tree that best approximates the underlying distribution of variables. Let $\mathcal{R}$ be the set of possible 1-order dependence trees. Formally, the DAG $R^{*}$ is a CLT if
\[
R^{*}\in\arg\min_{R\in\mathcal{\mathcal{R}}}D_{KL}(\nu||\nu_{R}),
\]
that is, $\nu_{R}$ minimizes Kullback-Leibler divergence, $D_{KL}$, from $\nu$ within that
set, where 
\[
D_{KL}(p||q)=\sum_{x}p(x)\log\left(\frac{p(x)}{q(x)}\right)
\]
when $X$ is discrete or
\[
D_{KL}(P||Q)=\int\log\left(\frac{f(x)}{g(x)}\right)P(dx)
\]
when $X$ is continuous, $Q\ll P$, $f$ is pdf of $P$, and~$g$
is the pdf of $Q$. 

\cite{chow1968approximating} show that the optimal tree can be found
using a greedy algorithm in polynomial time. The algorithm finds the minimum spanning tree for  appropriate weights. It considers undirected edges between each pair of variables and calculates
the mutual information between the variables related by each edge. Then, it sorts the edges  in decreasing order of informativeness. Starting with a graph
consisting of the first two edges, it checks whether adding the next edge
would create a cycle. If not, then it adds the edge to the graph. Otherwise,
it discards the edge. It proceeds to the next edge and continues until each edge is either added or discarded. The resulting graph can be oriented
to form a $1$-order dependence tree, and this tree has the lowest
divergence amongst all such trees.

\subsection{Normal variables and CLT}

With normally distributed variables, the relative quality of the approximation can be determined by
pairwise correlations.
\begin{lem}
\label{lem:KLD normal}
For any normal distribution $p$, there exists a constant $\kappa>0$
so that 
\begin{equation}
D_{KL}\left(P||P_{R}\right)=\kappa+\sum_{(i,j)\in R}\frac{1}{2}\ln\left(1-\rho\left(x_{i},x_{j}\right){}^{2}\right)\label{eq:KLD normal}
\end{equation}
for every 1-order dependence tree $R$.
\end{lem}
The divergence minimizing tree is the one that maximizies correlations
between linked variables. To illustrate with 3 normal variables, every
1-order dependence tree takes form $R=i\rightarrow j\rightarrow k$.
Then, the divergence can be written 
\begin{align*}
D_{KL}(P||P_{R}) & =\frac{1}{2}\ln\left[\left(1-\rho(i,j)^{2}\right)\sigma_{i}^{2}\sigma_{j}^{2}\right]+\frac{1}{2}\ln\left[\left(1-\rho(j,k)^{2}\right)\sigma_{k}^{2}\sigma_{j}^{2}\right]\\
 & \qquad+\frac{1}{2}\ln\left[\sigma_{j}^{2}\right]-\frac{3}{2}\ln\left(2\pi e\right)-h(p)\\
 & =\frac{1}{2}\ln\left(1-\rho(i,j)^{2}\right)+\frac{1}{2}\ln\left(1-\rho(j,k)^{2}\right)\\
 & \qquad-\ln\sigma_{1}-\ln\sigma_{2}-\ln\sigma_{3}+\frac{3}{2}\ln\left(2\pi e\right)-h(p)
\end{align*}
where $h(p)$ is the Shannon entropy of the distribution $p$. Here,
$\kappa=-\ln\sigma_{1}-\ln\sigma_{2}-\ln\sigma_{3}+\frac{3}{2}\ln\left(2\pi e\right)-h(p)$.
Inspecting the formula reveals that the CLT links every pair of variables
except the pair with minimal correlation.

\begin{figure}
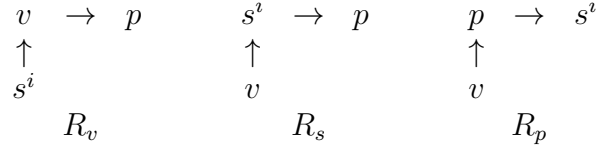

\begin{tabular}{ccc}
$v$ & $\rightarrow$ & $p$\\
$\uparrow$ 	&	&\\
$s^{i}$	&	&\\
& $R_v$ &
\end{tabular}
\qquad
\begin{tabular}{ccc}
$s^{i}$ &$ \rightarrow$ &$ p$\\
$\uparrow$\\
$v$\\
& $R_s$ &
\end{tabular}
\qquad
\begin{tabular}{ccc}
$ p$&$ \rightarrow$ &$s^{i}$ \\
$\uparrow$\\
$v$\\
& $R_p$ &
\end{tabular}
\caption{Possible $1$-order trees with $3$ variables}
\label{fig: CLT J1}
\end{figure}
With the three variables $v,s^{i},p$, there are three possible $1$-order
dependence trees (Figure \ref{fig: CLT J1}).%
\footnote{Since there are no colliders, any DAGs with the same undirected version generate the same approximation by \cite{Verma1991equivalence}.} Suppose agent tries to predict $v$. On the one hand, $R_{v}$ leaves
out the correlation between $p$ and $s^{i}$ conditional on $v$,
so beliefs about $v$ overreact to $s$ given $p$. However, it is
\emph{precise}, in that it results in beliefs about $v$ with relatively
low variance. On the other hand, $R_{s}$ (and $R_{p}$, with $p$
replacing $s$ in what follows) leaves out the correlation between
$v$ and $p$. Therefore, beliefs underreact to $p$ given $s$
relative to a Bayesian. However, beliefs are \emph{accurate}, in the
sense $E_{R_{S}}[v|s,p]=E[v|s]$.

\subsection{Higher order trees}
Increasing the order corresponds to an increase in
the depth of the network. This increases the quality of the approximation
at the cost of decreasing the amount of dimension reduction. This
captures the inherent tradeoff in machine learning: higher predictive
power requires more computation and entails a danger of overfitting. For textbook treatments of these topics, see Chapter 5 of \cite{hajek1992} or Chapter 6 of \cite{koski2009}.

In general, the determination of how large $K$ should be involves
tradeoffs. First, finding best $K$-order tree is NP-hard for $K\geq2$
(\cite{chickering1996learning}). Therefore, actually calculating
the optimum may not be feasible and one may have to satisfice instead.
Second, with limited (though still large) numbers of observation,
the ideal $K$ is determined by the bias-variance tradeoff in a process
called regularization. A smaller $K$ suffers from the potential for
bias due to omitted relationships. However, larger $K$ leads to more
variance in estimates. This is due to the risk of overfitting because
of outliers. In particular, the model must be estimated on each $K+1$
tuple of variables, and for a fixed dataset size, every particular
realization of these variables occurs less often. In general, algorithms trade-off the greater expressive
power of deeper learning (i.e., larger $K$) with the stability of
estimates by using penalty functions. Our use of CLT amounts to assuming
that the dataset is such that the penalty for $K>1$ is  large enough
that the algorithm chooses $K=1$ as optimal. \cite{spiegler2026machine} explicitly models this tradeoff in a  trust game.

\section{Asset market model}
\label{sec:Asset market}
There is a mass $I\in\{1,2,3,\dots\}$ of agents indexed by $i^{\prime}\in(0,I]$.
Trader $i^{\prime}\in(i-1,i]$ is referred to as a trader of type
$i$. Traders form a portfolio from a risky asset and a safe asset,
and are able to take arbitrarily large short and long positions provided
that they satisfy the budget constraint. The risky asset has a common
value $v$ where $v\sim N(0,\sigma_{v}^{2})$. The safe asset has
a return normalized to unity. Trader $i^{\prime}\in(i-1,i]$ maximizes
CARA utility with risk aversion parameter $\frac{1}{r^{i}}>0$, so
if she purchases $x$ units of the risky asset at price $p$, her
utility is $-\exp\left(-\frac{1}{r^{i}}x(v-p)\right)$

There are $I$ signals, one for each trader type, and all traders of
type $i$ observe $s^{i}\in\mathbb{S}^{i}=\mathbb{R}^{J}$. Each
$s^{i}$ is a $J$ dimensional vector with $s_{j}^{i}=v+\epsilon_{j}^{i}$,
where $\epsilon^{i}\sim N(0,\Sigma^{i})$ for $i=1,\dots,I$, where
each $\Sigma^{i}$ is a symmetric, positive definite $J\times J$
matrix, independently of all other variables. Trader $i^{\prime}\in(i-1,i]$
observes the price $p$ of the risky asset and $s^{i}$. Let $\sigma_{i,j}^{2}=\Sigma_{jj}^{i}$
and $\vec{s}=(s^{1},\dots,s^{I})$. Supply of the asset equals $u$,
where $u\sim N(0,\sigma_{u}^{2})$ with $\sigma_{u}^{2}>0$ and is
independent of all other variables.

Equilibrium consists of three endogenous objects. The first two are
standard: a mapping from fundamentals to the price and a demand function
for each trader that clears the market. The remaining object is a
CLT for each trader, which we describe via a distribution over $1$-order
dependence trees for each type $i$. We consider linear equilibria
where the price $p$ is an affine function of the vector $\left(s,u\right)=\left(s^{1},\dots,s^{I},u\right)$,
i.e., there is an $\alpha\in\mathbb{R}^{IJ+2}$ so that $p=\alpha\cdot(1,\vec{s},u)$.
For such an $\alpha$, denote by $\nu(\cdot|\alpha)$ the distribution
over all variables that results, and for each $i$, $\nu^{i}(\cdot|\alpha)$
the distribution over $(v,s_{1}^{i},\dots,s_{J}^{i},p)$. 
Denote the $1$-order dependence trees over $N^{i}=\left\{ s_{1}^{i},\dots,s_{J}^{i},v,p\right\} $ including $\{v,x\}$ for at least one $x\in N^{i}\setminus\{v\}$ by $\mathcal{R}^{i}$. 

The key difference from \cite{hellwig1980aggregation} is that traders update beliefs using a CLT rather than Bayes rule. The algorithm determines the tree $R\in \mathcal{R}^{i}$ that best fits the equlibrium distribution of feedback. Crucially, this calculation happens ex ante, so no updating takes place prior to choosing the tree. Then, the approximation $\nu^i_R(\cdot |\alpha)$ according to $R$ is updated to $\nu^i_R(\cdot|\alpha,s^i,p)$, and the trader's demand maximizes expected utility according to this belief. This captures a trader constrained to be unable to update $\nu^i(\cdot|\alpha)$ unaided. The true distribution $\nu^i(\cdot|\alpha)$ is typically not equal to $\nu^i_R(\cdot|\alpha)$ for any $R\in \mathcal{R}^i$, so the model is misspecified. 
\begin{defn}
\emph{A CLT equilibrium} is an $\alpha\in\mathbb{R}^{IJ+2}$
and a vector $\vec{\mu}=\left(\mu^{1},\dots,\mu^{I}\right)$ where
$\mu^{i}$ is a distribution over $\mathcal{R}^{i}$ so that:
\begin{enumerate}
\item for every $u\in\mathbb{R}$ and $\vec{s}\in\mathbb{R}^{IJ}$, $p=\alpha\cdot(1,\vec{s},u)$; 
\item for each trader $i$ and tree $R\in\mathcal{R}^{i}$, every price
$p\in\mathbb{R}$, and all signals $s^{i}\in\mathbb{R}^{J}$,
\[
x_{R}^{i}\left(s^{i},p\right)=\arg\max_{x}-\int_{\mathbb{R}}\exp\left(-\frac{1}{r^{i}}x\left(v-p\right)\right)\nu_{R}\left(dv|\alpha,s^{i},p\right);
\]
\item each trader uses a CLT: for every $i$,
\[
\mu^{i}(R)>0\implies R\in\arg\min_{R\in\mathcal{R}^{i}}D_{KL}\left(\nu^{i}(\cdot|\alpha)||\nu_{R}^{i}(\cdot|\alpha)\right);
\]
\item and the distribution over trees clear the market for every $(\vec{s},u)\in\mathbb{R}^{IJ+1}$
\[
\sum_{i=1}^{I}\sum_{R\in\mathcal{R}^{i}}\mu^{i}(R)x_{R}^{i}(s^{i},p)=u.
\]
\end{enumerate}
\end{defn}
First, the price is an affine function of the vector of signals and
noise trader demand. Second, each trader maximizes expected utility
given her tree, private information $s^{i}$, and the market price.
Consequently, the demand from a trader of type $i$ who uses the tree
$R$ is
\begin{equation}
x_{R}^{i}(s^{i},p)  =r^{i}\frac{E_{R}[v|s^{i},p]-p}{Var_{R}[v|s^{i},p]}
\label{eq:CLT demand}
\end{equation}
following the usual formulas for updating a normal distribution. Third,  the distribution of trees $\mu^{i}$ attaches support only to
trees that minimize divergence given the equilibrium distribution. 
Finally, given the demand functions above and the distribution of trees
$\vec{\mu}$, the market clears for every realization of signals and
noise trader demand.  Aggregate demand from traders of type $i$ is 
\[
\sum_{R\in\mathcal{R}^{i}}\mu^{i}(R)x_{R}^{i}(s^{i},p).
\]
Market aggregate demand sums the above across types, and must equal $u$, the aggregate supply of the asset. 

We interpret equilibrium as follows.
Each trader observes a large dataset drawn from the equilibrium distribution. They input this dataset into the CLT algorithm, along with their current observations. The algorithm outputs a predicted distribution of $v$ according to a tree that is one of the best approximations of the dataset. The algorithm's choice of tree optimal according to a statistical criterion as in \cite{jehiel2026endogenous}, so the predictions need not yield the most expected utility (see Section \ref{sec:benchmarks}).  As in \cite{jehiel2026endogenous}, the equilibrium may require the algorithm to randomize between the different outputs of the algorithm, which could be interpreted as reflecting differences in the datasets across traders of the same type. Equilibrium can be though of as a steady state of this process repeated over many periods. The distribution of trees and prices both clears the market in this period and generates the dataset inputted into the algorithm in the next period.

\section{Analysis}
\label{sec:analysis}
We begin by showing that an equilibrium exists for every economy as
above. Then, we turn to the properties of this equilibrium. In every
equilibrium of every economy, even approximate informational efficiency
is impossible. 

\subsection{Existence}

First, we show that an equilibrium exists.
\begin{thm}
\label{thm: CLT exists}
A CLT equilibrium exists for any collection of parameters with $\sigma_u^2>0$.
\end{thm}
Given a fixed distribution of models $\mu$, there exists a unique
price that clears the market. The relative weights on different signals
are determined by the ratio 
\[
\frac{\alpha_{i,j}}{\alpha_{k,l}}=\frac{\mu^{i}\left(\left\{ R:\{v,s_{j}^{i}\}\in R\right\} \right)\sigma_{i,j}^{-2}r^{i}}{\mu^{k}\left(\left\{ R:\{v,s_{l}^{k}\}\in R\right\} \right)\sigma_{k,l}^{-2}r^{k}}.
\]
Signals receive higher weights when they are more frequently used,
more precise, or utilized by more risk-tolerant traders. Notably, and
unlike in standard Bayesian frameworks, the correlation between the
signal and the price does not influence these weights. The resulting
price determines the correlation between the fundamental value $v$
and the price $p$, as well as between $p$ and each signal $s_{j}^{i}$.
The distribution of trees adjusts endogenously so that only the best
approximations are used. The creates feedback effects on pricing and
correlations. 

The equilibrium balances two considerations. The more that the price
responds to a signal, the higher the correlation between the two. Therefore,
more traders' trees link that signal to the price, and fewer link it to the value. Consequently, demand becomes less responsive
to changes in the signal. Conversely, the less that the price responds
to a given signal, the lower the correlation between the two. Consequently,
more traders' trees link it to the value, and so demand becomes more
responsive to it. In equilibrium, the price must respond just enough
so that the market clears. If the correlation with price is too large
for every signal, then all traders rely exclusively on information
from the price. If it is instead too low for enough signals, then
all traders use only their private information. Either poses difficulty
for market clearing.

We prove existence of equilibrium by solving a nested fixed point problem.
First, we characterize the unique price that clears the market for
a given $\vec{\mu}$ via a continuous function, $\alpha\left(\vec{\mu}\right)$.
Then, we define a correspondence from a vector of distributions over
trees to itself. The correspondence evaluated at $\vec{\mu}$ puts
weight only on the trees that minimize divergence given the price
is generated by $\alpha\left(\vec{\mu}\right)$ . Any fixed point
of the correspondence is an equilibrium.

\subsection{Equilibrium information}

Replacing $s^{i}$ with $\hat{s}^{i}=1^{\prime}\left(\Sigma^{i}\right)^{-1}s^{i}$,
our setup is identical to that of \cite{hellwig1980aggregation}.
In that model, all traders are Bayesian and there is a unique equilibrium
price $p$ that reflects all private signals. In particular, the equilibrium
price is \emph{approximately informationally efficient}, in the following
sense. As noise $\sigma_{u}$ goes to zero, 
\[
E[v|p]\rightarrow E[v|\vec{s}].
\]
That is, beliefs approach what they would be if all private signals
were public.
Given the normal distribution, this is equivalent to
\[
\rho(v,p) \rightarrow\rho(E[v|\vec{s}],v) \equiv \rho^*.
\]
The limiting price is a sufficient statistic
for signals and so maximizes the correlation between the price and
value.

In our setting, approximate informational efficiency fails.
\begin{thm}
\label{thm: info efficiency}
Fixing $I>1$, $\sigma_{v}$, and $\Sigma^{i}$ for all $i$, let
$\rho^{*}=\rho(E[v|\vec{s}],v) $. There exists $\delta>0$ so
that for any $\sigma_{u}>0$, $\rho(p,v)^{2}<\left(\rho^{*} \right)^2-\delta$ in
any CLT equilibrium.
\end{thm}
The result shows that not all private information is aggregated in
the market price, even approximately. After trading, private information
still has value to a Bayesian decision maker. Moreover, traders get
at least as much uncertainty reduction from using the tree as a Bayesian
would from the price. Put differently, traders perceive a non-negative
and sometimes strictly positive value of using the algorithm and private
information relative to just the public price. Since training a machine
learning algorithm is typically costly, this result is consistent
with persistent use of them in a market.

The proof can be found in the appendix. The key step utilizes the
data processing inequality. Recall that $\mathbb{E}[v|\vec{s}]$ is
a sufficient statistic for all private information, and that for pairwise normally distributed variables, mutual information is an increasing function of correlation squared. The data-processing
inequality \citep[Theorem 2.8.1 of][]{CoverThomas2006} says that if $X\rightarrow Y \rightarrow Z$ is a Markov chain, then the mutual information between $X$ and $Y$ exceeds that between $X$ and $Z$, with equality only if $X \rightarrow Z \rightarrow Y$ is also a Markov chain.  Therefore, 
$\rho(\mathbb{E}[v|\vec{s}],s_{j}^{i})>\rho(v,s_{j}^{i})$ and $\rho(v,\mathbb{E}[v|\vec{s}])>\rho(v,s_{j}^{i})$.
If the correlation between price and $v$ is sufficiently close to
$\rho(v,\mathbb{E}[v|\vec{s}])$, then $\rho(p,s_{j}^{i})>\rho(v,s_{j}^{i})$
and $\rho(v,p)>\rho(v,s_{j}^{i})$ as well. But then by Lemma 1, the
unique CLT is 
\[
\begin{array}{cccc}
v\leftarrow & p&\rightarrow & s_{1}^{i}\\
 & \downarrow&\searrow&\\
 & s_{J}^{i} & &\cdots
\end{array}
\]
for every type of trader, so demand is invariant to private signals
given the price. But then for markets to clear, $p=\alpha_{u}u$,
implying $\rho(p,v)=0$ and $\rho(p,s_{j}^{i})=0$, a contradiction. 

\section{Examples}
\label{sec:examples}
To understand the forces in the model, we consider several example
economies. The first considers information aggregation in the classic
\cite{hellwig1980aggregation} setup with identical precisions and risk tolerance across traders. Then, the algorithm trades off between public information $p$ and single-dimensional private information $s^{i}$.
Our focus is on how the equilibrium differs from Bayesian benchmarks. The second
focuses on what aspects of the information get used by the tree in
a setting where all information is public. Then, the algorithm trades
off between different dimensions of the information and potentially
substitutes some of them with the garbling  $p$. The general case has features of
both, and the two are informative of what the general model looks
like.

\subsection{Aggregation across signals}

First, suppose that all traders receive a one-dimensional signal ($J=1$),
that they all have the same risk aversion ($r^{i}=r$ for all $i$),
and that signals are equally precise ($\Sigma_{\epsilon}^{i}=\left[\sigma_{\varepsilon}^{2}\right]$).
Denote by $R_{j}$ for $j\in\{v,s,p\}$ the CLT that has two edges
involving the node $j$. This is the model of \cite{hellwig1980aggregation},
specialized so that agents are symmetric.
\begin{thm}
\label{thm: CLT J=1}
If $I>1$, $J=1$, $r^{i}=r$ for all $i$, and $\Sigma_{\epsilon}^{i}=\left[\sigma_{\varepsilon}^{2}\right]$
for all $i$, there is a symmetric CLT
equilibrium $\left(\alpha,\vec{\mu}\right)$. If $\left(I^{2}-2I\right)\sigma_{v}^{2}>\sigma_{\varepsilon}^{2}$,
then for every $i$,
\[
\mu^{i}(R_{v})=\min\left\{ 1,\sqrt{\frac{\sigma_{\varepsilon}^{2}\sigma_{u}^{2}\sigma_{v}^{2}}{r^{2}\left(\sigma_{\varepsilon}^{2}+I \sigma_{v}^{2}\right)}}\right\} =1-\mu^{i}(R_{p}).
\]
 If $\left(I^{2}-2I\right)\sigma_{v}^{2}<\sigma_{\varepsilon}^{2}$,
then for every $i$
\[
\mu^{i}(R_{s})=\min\left\{ 1,\sqrt{\frac{ \sigma_{u}^{2}\sigma_{\varepsilon}^{2}}{r^{2}I \left(I-1\right)}}\right\} =1-\mu^{i}(R_{p}).
\]
In the former case, $\rho(p,v)\geq\frac{I\sigma_{v}^{2}}{I\sigma_{v}^{2}+\sigma_{\varepsilon}^{2}}$and
in the latter $\rho(p,v)\geq\frac{\sigma_{v}}{\sqrt{\sigma_{v}^{2}+\sigma_{\varepsilon}^{2}}}$,
with equality whenever $\mu^{i}(R_{p})>0$.
\end{thm}
The result provides an explicit characterization of the equilibrium.
The two cases obtain because for any $\alpha$, $$\left(I^{2}-2I\right)\sigma_{v}^{2}\geq\sigma_{\varepsilon}^{2} \iff \rho(v,p)\geq\rho(p,s^{i}),$$ so the inequality determines whether the CLT can contains an edge from $v$ to both $p$ and $s^{i}$
or not. When $\sigma_{u}^{2}$ is large, the equilibrium price is
not very informative about $v$ nor highly correlated with any other
variables. Consequently, the algorithm selects either $R_{s}$ or
$R_{v}$ for all traders. When $\sigma_{u}^{2}$ is small enough,
the price is more correlated with $v$. If the algorithm never selected
the tree $R_{p}$, then the price would be sufficiently strongly correlated
with both $s^{i}$ and $v$ that $R_{p}$ would be the unique CLT,
a contradiction. Consequently, the distribution of trees and price
must adjust so that the algorithm finds both trees optimal.

\subsubsection{Benchmark:  \cite{hellwig1980aggregation} }
Consider first the economy where all traders are rational. Denote
the equilibrium price $p^{h}=\alpha_{s}^{h}I^{-1}\sum s^{i}+\alpha_{u}^{h}u$.
This special case of \cite{hellwig1980aggregation} is explicitly
solved in Theorem 6.1 of \cite{kyle1989informed} via the system of equations
\begin{align*}
\frac{\varphi}{\left(1-\varphi\right)^{3}} & =r^{2}\frac{I-1}{\sigma_{u}^{2}\sigma_{\varepsilon}^{2}}\\
\alpha_{s}^{h} & =\frac{\sigma_{\varepsilon}^{-2}\left(1+\varphi\left(I-1\right)\right)}{\sigma_{v}^{-2}+\sigma_{\varepsilon}^{-2}\left(1+\varphi\left(I-1\right)\right)}\\
\alpha_{u}^{h} & =\alpha_{s}^{h}\sigma_{\varepsilon}^{2}\frac{1}{Ir(1-\varphi)}.
\end{align*}
Letting $\tau= \sigma_v^{-2}+(1+\varphi (I-1))\sigma_{\varepsilon}^{-2}$, 
equilibrium demand is \[ 
x^i(p,s^i)=r \mathbb{E} \left[ v|s^i,p\right] \tau =r(1-\varphi)\sigma_{\varepsilon}^{-2}s^i-r\left(\tau-I \varphi \frac{\sigma_{\varepsilon}^{-2}}{\alpha_s} \right)p.
\]
One can then verify that then $p^g$ clears the markets state-by-state.

\subsubsection{Benchmark: \cite[][ henceforth, GS]{grossman1980impossibility}}
We introduce a slight extension of GS to dispersed information. A trader $i' \in (i-1,i] $ can either give up $c>0$ units of the safe asset to see the signal $s^i$ in addition to the price or  submit a demand that is only a function of the price. 

Let $x^i_U(p)$ and $x^i_I(s^i,p)$ be the demands maximize the utility of a type $i$ trader given $p$ and $(s^i,p)$ are observed, respectively. Equilibrium consists of a price $p^{g}=\alpha_{s}^{g}I^{-1}\sum_i s^{i}+\alpha_{u}^{g}u$ and a $\lambda\in [0,1]$ fraction of traders that acquire the signal so that
\[
\mathbb{E}\left[-\exp \left( - x_U^i (p)(v-p)/r \right) \right] \leq \mathbb{E} \left[-\exp\left( -\left( x_I^i (s^i,p)(v-p) -c \right)/r
\right) \right],
\]
with equality whenever $\lambda\in (0,1)$,
and so that \[
\sum_i [(1-\lambda) x_U^i (p)+\lambda x_I^i (s^i,p)] = u.
\]
Following \cite{kyle1989informed}, define
\begin{align*}
\varphi^{I} & =\frac{\left(\alpha_s^{g}\right)^{2} \left(I-1\right)\sigma_{\varepsilon}^{2}}{\left(I-1\right)\left(\alpha_s^{g}\right)^{2} \sigma_{\varepsilon}^{2}+I^{2}\left(\alpha_u^{g}\right)^{2}\sigma_{u}^{2}}\\
\varphi^{U} & =\frac{\left(\alpha_s^{g}\right)^{2} I\sigma_{\varepsilon}^{2}}{\left(\alpha_s^{g}\right)^{2} I\sigma_{\varepsilon}^{2}+I^{2}\left(\alpha_u^{g}\right)^{2}\sigma_{u}^{2}}\\
\tau^I&= \sigma_v^{-2}+(1+\varphi^I (I-1))\sigma_{\varepsilon}^{-2}\\
\tau^U &= \sigma_v^{-2}+I \varphi^U \sigma_{\varepsilon}^{-2}
\end{align*}
so that, using Bayes rule, we have
 \[ 
x^i_I(p,s^i)=r \mathbb{E} \left[ v-p|s^i,p\right] \tau^I =r(1-\varphi^I)\sigma_{\varepsilon}^{-2}s^i-r\left(\tau^I-I \varphi^I \frac{\sigma_{\varepsilon}^{-2}}{\alpha^g_s} \right)p.
\]
and \[ 
x^i_U(p)=r \mathbb{E} \left[ v-p|p\right] \tau^U =-r\left(\tau^U-I \varphi^U \frac{\sigma_{\varepsilon}^{-2}}{\alpha^g_s} \right)p.
\]
Calculating expected utility, the  indifference condition becomes
\begin{equation}
\label{eq:GS indif}
\sqrt{\frac{\tau^I}{\tau^U}}=\exp[c/r].
\end{equation}
as in GS, and this combined with the equations 
\begin{align*}
\left(I-1\right)\lambda^{2}r^{2} & =\sigma_{\varepsilon}^{2}\sigma_{u}^{2}\frac{\varphi^{I}}{(1-\varphi^{I})^{3}}\\
\alpha_{s}^{g} & =\sigma_{\varepsilon}^{-2}\frac{\lambda\left(1-\varphi^{I}\right)+\left(\lambda \varphi^I+(1-\lambda)\varphi^U \right)I}{\tau}\\
\alpha_{u}^{g} & =-\frac{\lambda\left(1-\varphi^{I}\right)+\left(\lambda \varphi^I+(1-\lambda)\varphi^U \right)I}{Ir\tau\lambda\left(1-\varphi^{I}\right)}.
\end{align*}
pins down the equilibrium for sufficiently low cost so that Equation (\ref{eq:GS indif}) holds with equality.

\subsubsection{Comparison}
\label{sec:benchmarks}

The CLT Equilibrium differs from that above \cite{hellwig1980aggregation}
in several ways. Denote $p^{CLT}$ the equilibrium price in a CLT equilibrium.  In particular, for sufficiently small $\sigma_{u}^{2}$,
 $p^{CLT}$ is both less informative and more volatile than $p^h$. 

\begin{cor}
For $\sigma_{u}^{2}$ small enough, 
$\rho(v,p^{CLT})<\rho(v,p^h)$ and  $Var(p^{CLT})>Var(p^h)$.
For $c$ small enough, 
$\rho(v,p^{CLT})<\rho(v,p^g)$ and $Var(p^{CLT})>Var(p^g)$.
\end{cor}

As $I$ goes to infinity, both CLT and Hellwig prices become perfectly correlated with the state. However, the CLT equilibrium price converges more slowly  than does the Hellwig equilibrium.
\begin{cor}
\label{cor: rate of convergence}
Let $p^{CLT}(I)$ and $p^{h}(I)$ be equilibrium prices in CLT and Hellwig models, respectively, as a function of the number of traders. Then, $\rho\left(v,\mathbb{E}[v|(s_i)_{i=1}^I] \right)^2-\rho\left(v,p^{CLT}(I)\right)^2=O(I^{-1})$ while $\rho\left(v,\mathbb{E}[v|(s_i)_{i=1}^I]\right)^2-\rho \left(v,p^{h}(I)\right)^2=O(I^{-4/3})$ 
\end{cor}
As $I$ gets large, the equilibrium price with both CLT and Bayesian traders becomes perfectly informative. However, the prices are differentially informative for any finite number of traders. In particular, the CLT equilibrium price approaches the limit at a slower rate. It thus remains less informative regardless of the number of traders.

\begin{figure}
\includegraphics[width=0.4\textwidth]{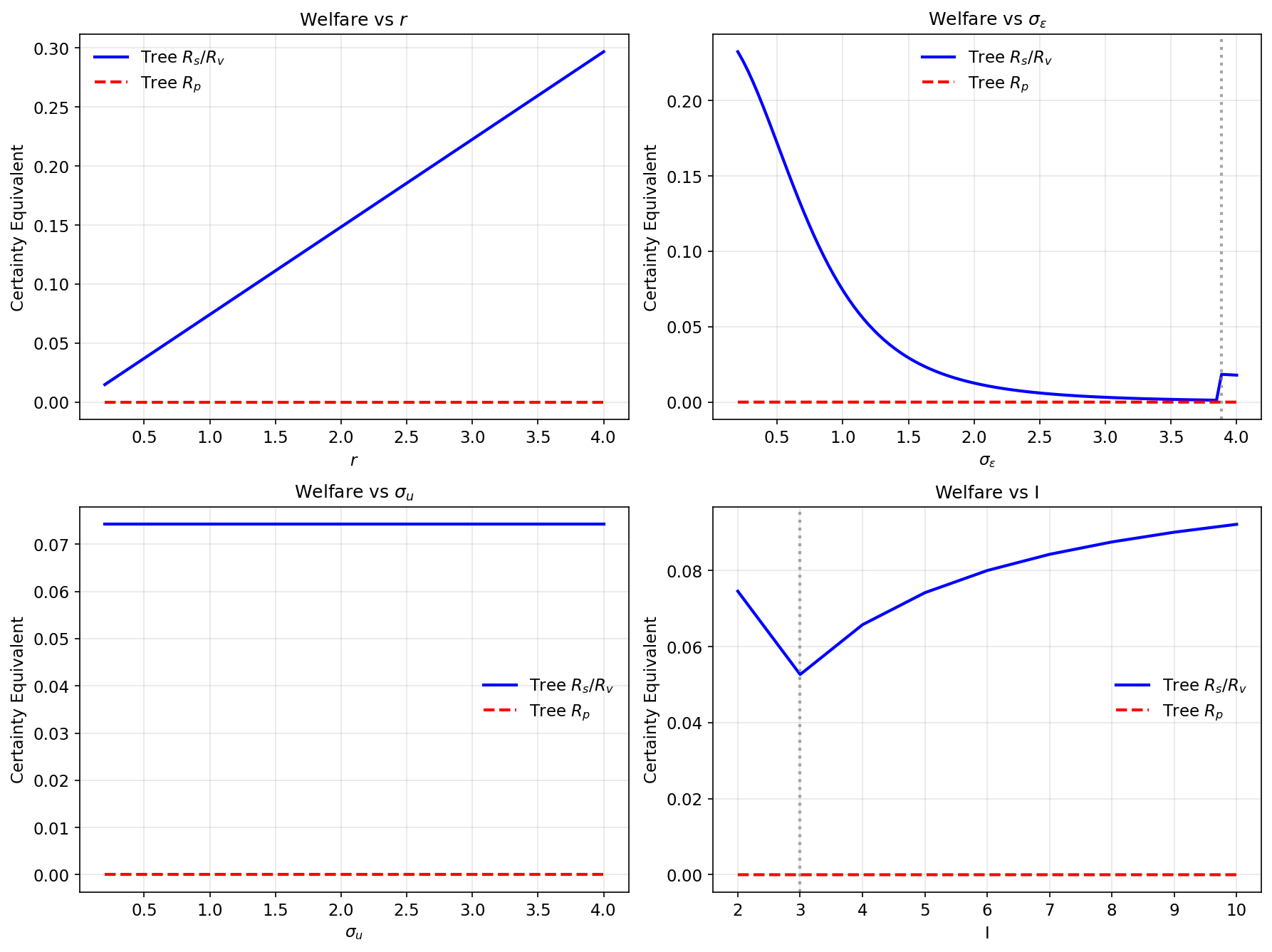} \qquad \includegraphics[width=0.4\textwidth]{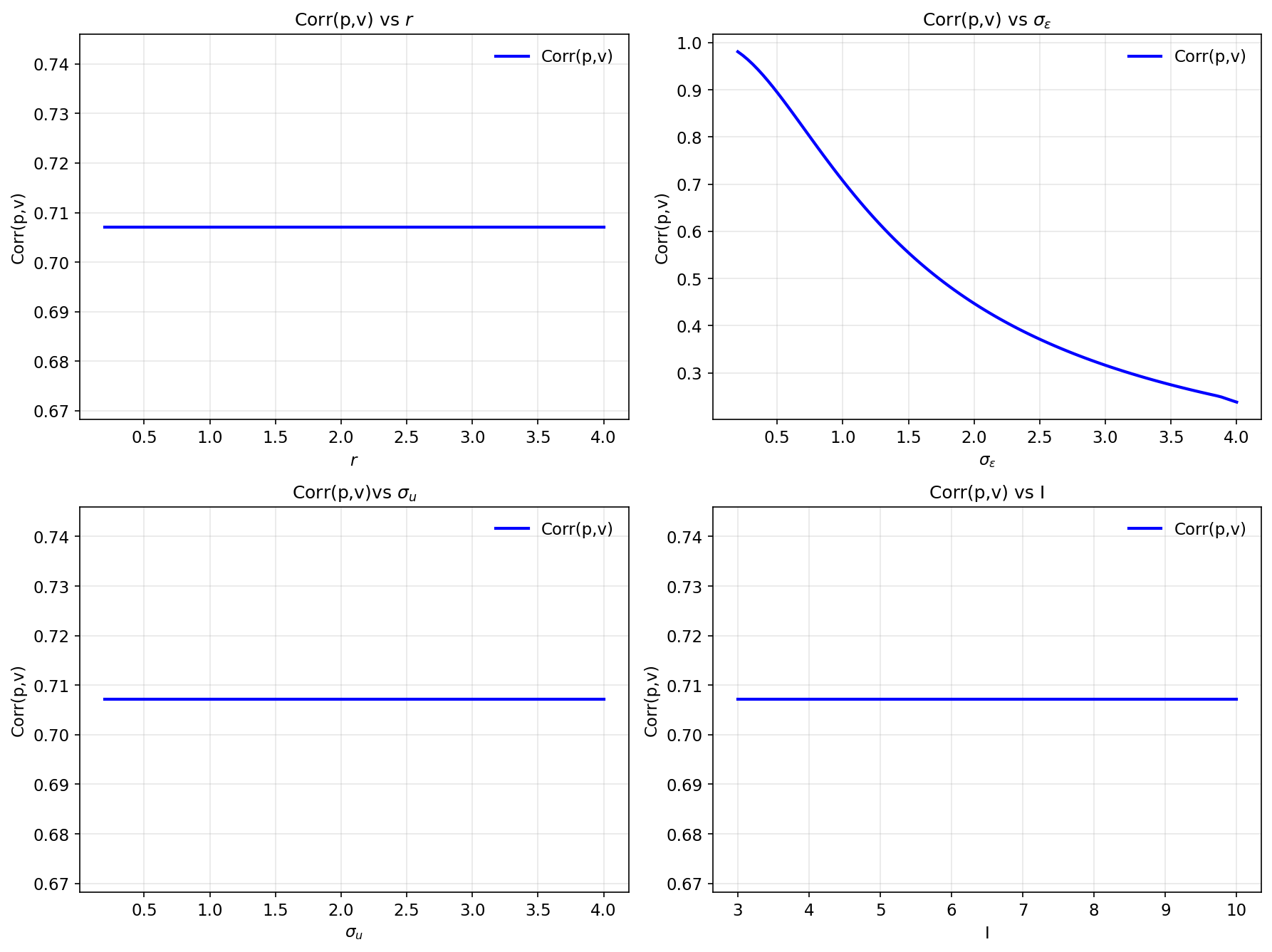}
\caption{Welfare and correlation in CLT equilibrium}
\label{fig:CLT welfare}
\end{figure}

\begin{figure}
\includegraphics[width=0.4\textwidth]{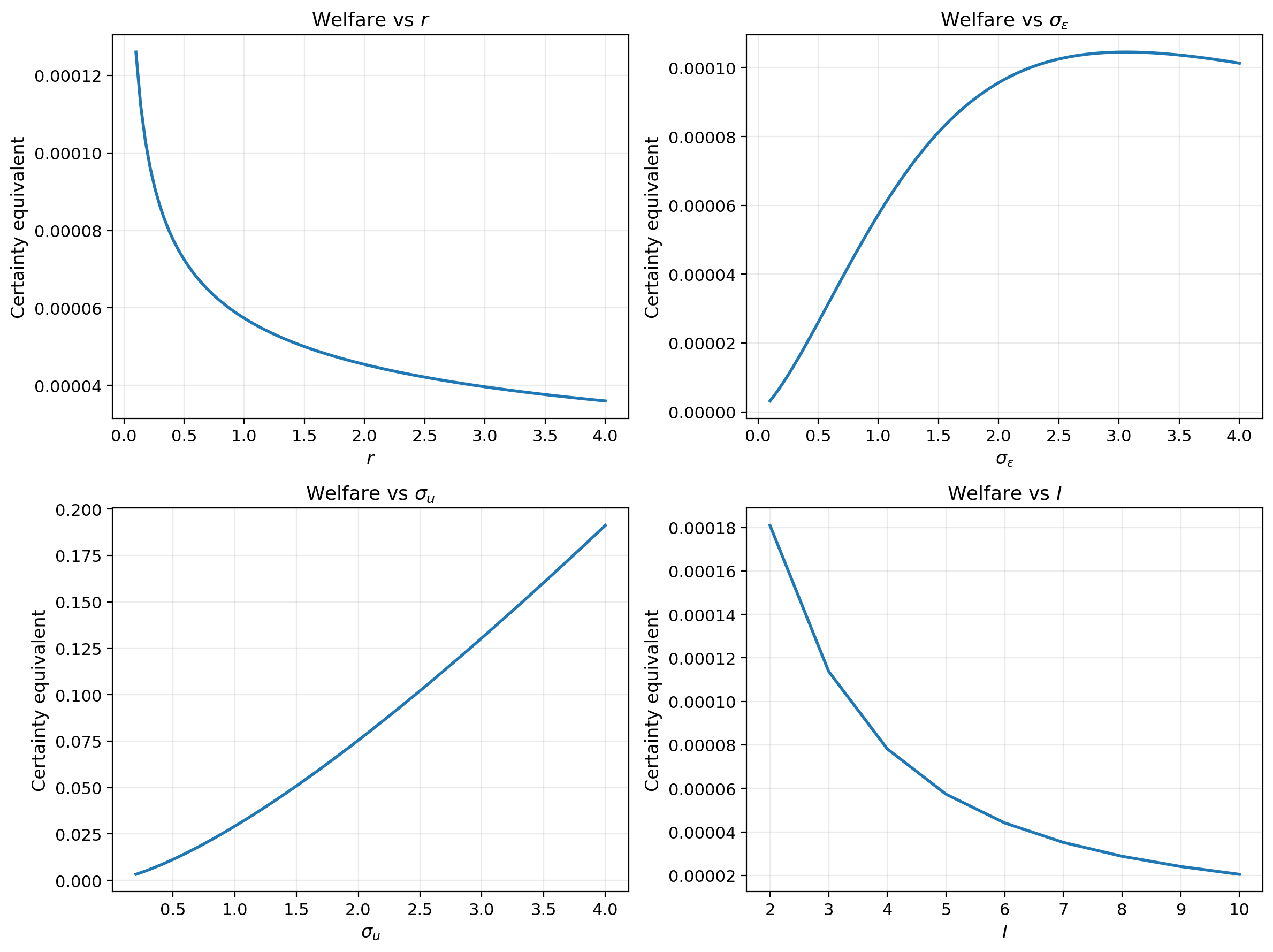}\qquad \includegraphics[width=0.4\textwidth]{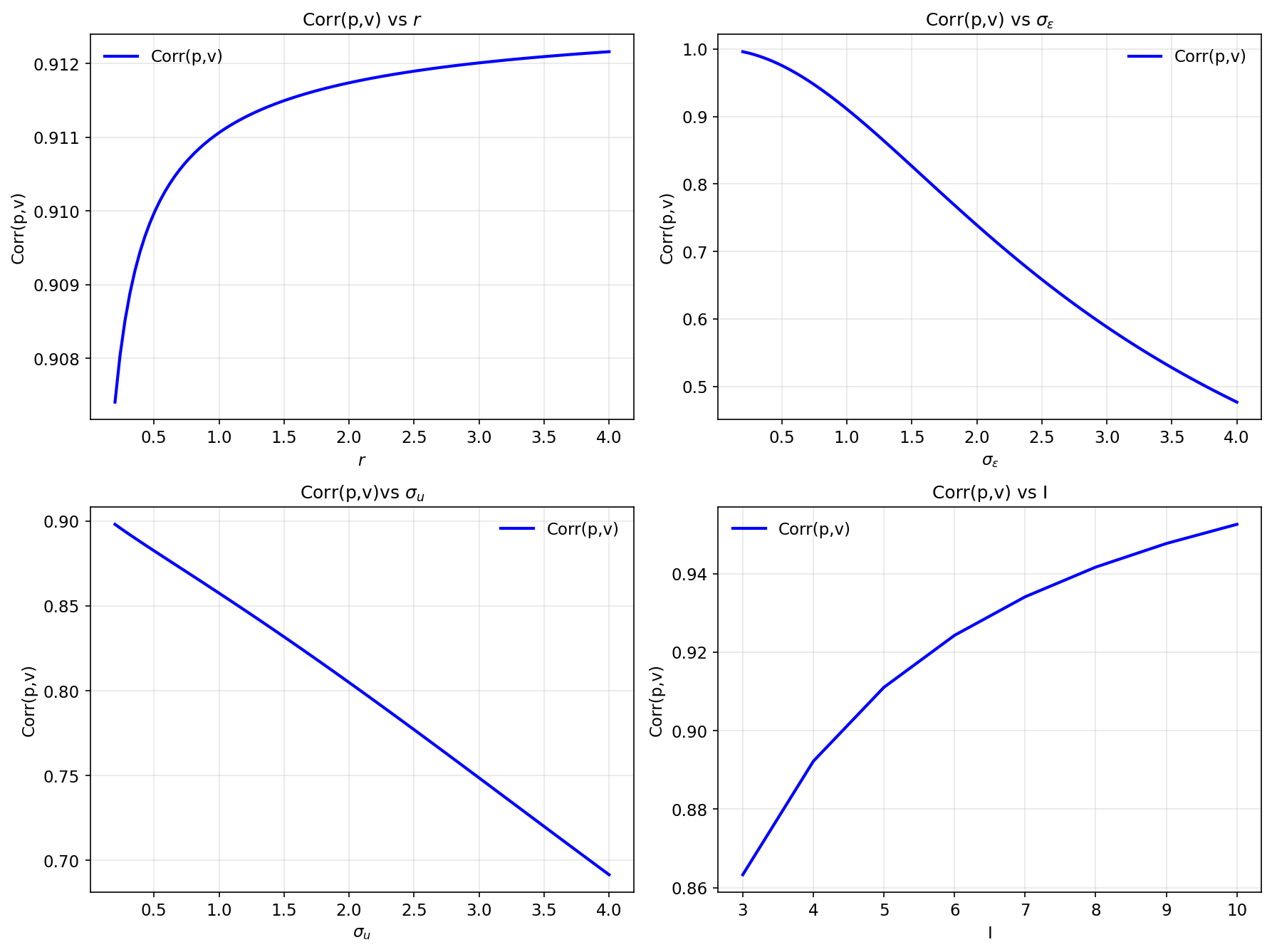}
\caption{Welfare and correlation in Hellwig equilibrium}
\label{fig:Hellwig welfare}
\end{figure}

\begin{figure}
\includegraphics[width=0.4\textwidth]{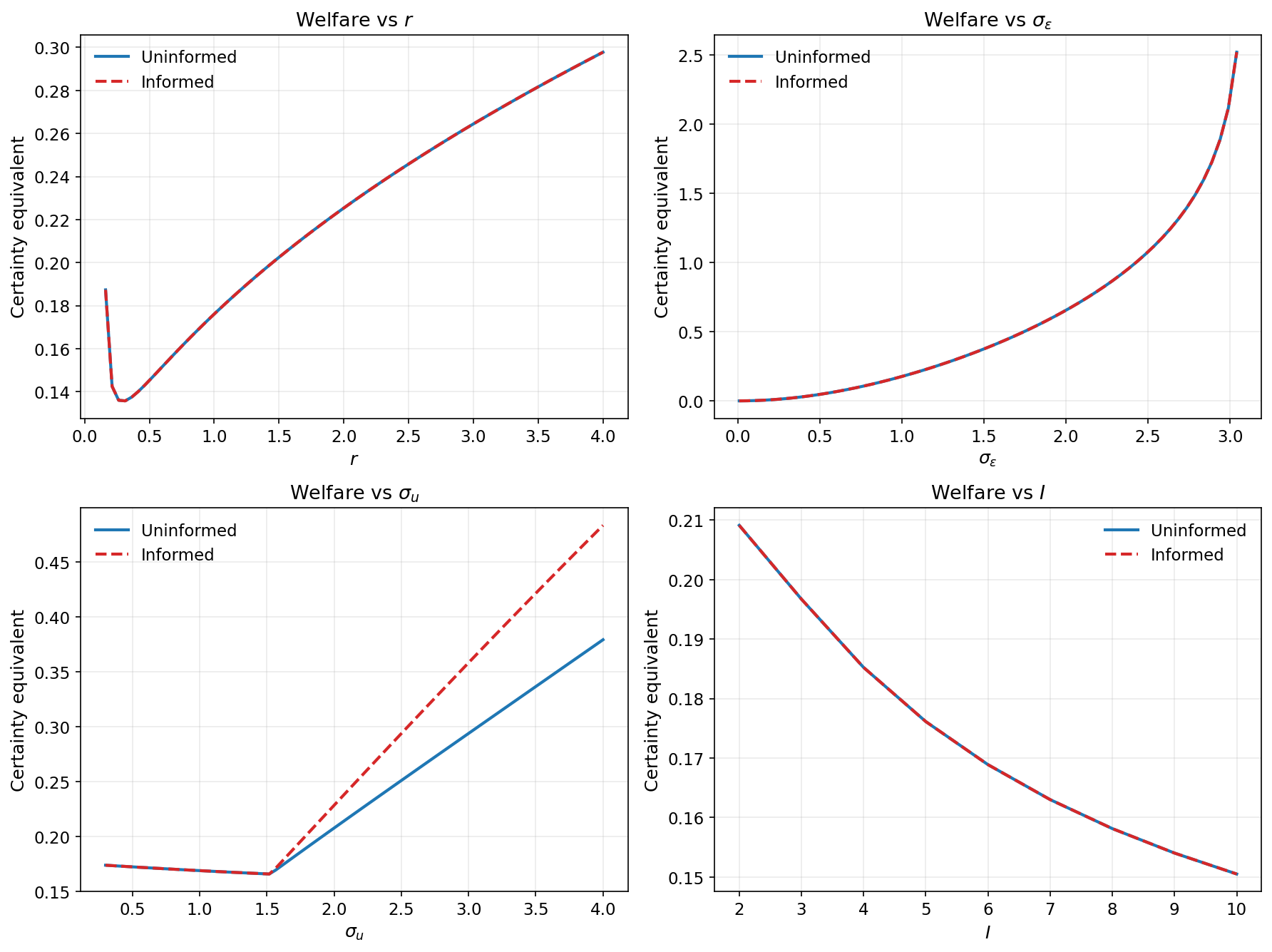}\qquad \includegraphics[width=0.4\textwidth]{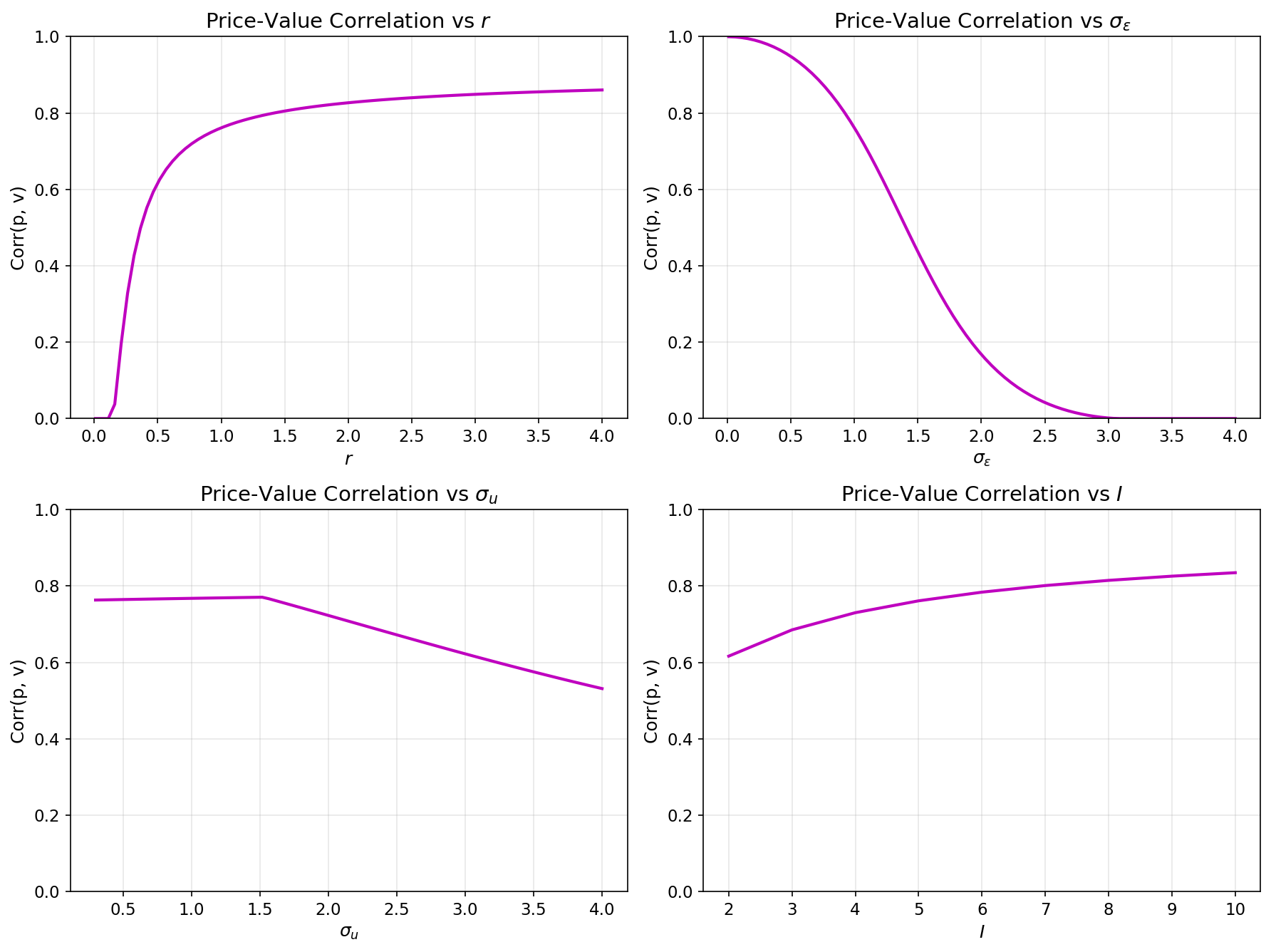}
\caption{Welfare and correlation in GS equilibrium}
\label{fig:GS welfare}
\end{figure}

The CLT equilibrium has significantly different comparative statics on both welfare and correlation between price and value than either of the two other equilibria. See Figures \ref{fig:CLT welfare}, \ref{fig:Hellwig welfare}, and \ref{fig:GS welfare}, which plot the equilibrium correlations between value and price and the certainty equivalents of each trader by tree. In CLT these equivalents are calculated with respect to the objective equilibrium distribution of price and the demand functions that result from the equilibrium distribution of trees. These take the baseline parameters $I=5$, $r=1$, $\sigma_v^2 =1$, $\sigma_\varepsilon^2=1$, and $\sigma_u=0.01$.

Notice that welfare depends on the tree that the algorithm selects, and trees that are equally good approximations may yield different welfare (according to the true distribution of price). In particular, ex ante expected utility $\mathbb{E}_{\nu} \left[-\exp \left(-r^{-1}x^i(p,s)(v-p) \right) \right]$ equals \[
\frac{-r}{ \sqrt{\left(r+Cov(x^i(p,s),v-p) \right)^2-  Var(x^i(p,s)) Var(v-p) }}.\]
Utility is increasing in how correlated demand is with $v-p$ and decreasing in its own variance and that of $p$. We can show, for instance, that in equilibrium the utility of a trader using $R_s$ is higher than that of one using $R_p$. For the former, $Var(x^i(p,s))$ is smaller and $Cov(x^i(p,s),v-p)$ is larger. The above plots use the certainty equivalent to adjust for risk aversion.

\subsection{Aggregation within signals}
\begin{figure}
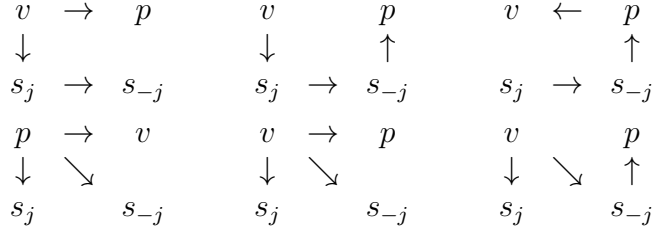

\begin{align*}
\begin{array}{ccc}
v & \rightarrow & p\\
\downarrow\\
s_{j} & \rightarrow & s_{-j}
\end{array}
\qquad
\begin{array}{ccc}
v &  & p\\
\downarrow &  & \uparrow\\
s_{j} & \rightarrow & s_{-j}
\end{array}\qquad\begin{array}{ccc}
v & \leftarrow & p\\
 &  & \uparrow\\
s_{j} & \rightarrow & s_{-j}
\end{array}\\
\begin{array}{ccc}
p & \rightarrow & v\\
\downarrow & \searrow\\
s_{j} &  & s_{-j}
\end{array}
\qquad
\begin{array}{ccc}
v & \rightarrow & p\\
\downarrow & \searrow\\
s_{j} &  & s_{-j}
\end{array}
\qquad
\begin{array}{ccc}
v &  & p\\
\downarrow & \searrow & \uparrow\\
s_{j} &  & s_{-j}
\end{array}
\end{align*}
\caption{Possible CLTs with $J=2$}
\label{fig:CLT J2}
\end{figure}

Our second main benchmark concerns high-dimensional public information. The simplest case is captured by  $I=1$, $J=2$, and 
\[
s^{i}|v\sim\mathcal{N}\left(v\vec{1},\left[\begin{array}{cc}
\sigma_{1}^{2} & \bar{\rho}\sigma_{1}\sigma_{2}\\
\bar{\rho}\sigma_{1}\sigma_{2} & \sigma_{2}^{2}
\end{array}\right]\right)
\]
where $\bar{\rho}\in(-1,1)$. Without loss, assume that $\sigma_{1}^{2}\leq\sigma_{2}^{2}$ so $\rho(v,s_{1}) \geq \rho(v,s_{2})$. There are more trees than
before. Figure (\ref{fig:CLT J2})  contains some  of them.

\begin{thm}\label{thm: specialization}
There exists $\hat{\rho}\left(\sigma_{v},\sigma_{1},\sigma_{2}\right)>0$ so that  the CLT equilibrium  $\mu$ has  $$\mu\left(\left\{ R:\left\{ \left\{ v,s_{1}\right\} ,\left\{ v,s_{2}\right\} \right\} \subset R\right\} \right)>0$$
only if $\bar{\rho}\leq \hat{\rho}\left(\sigma_{v},\sigma_{1},\sigma_{2}\right)$.
\end{thm}

Even when all traders observe the same signals, equilibrium features specialization when $\bar{\rho}$ is large. Only when the correlation between the two dimensions, conditional on $v$, is sufficiently low does any trader focus on both dimensions at the same time.  When signals are sufficiently correlated with each other, the algorithm faces a tradeoff: each of the two signals may be both strongly correlated with the value but they are also strongly correlated with each other.  The optimal tree trades off more precise beliefs by having more edges with more accuracy in not double counting some correlated information. For certain parameters, the algorithm may pick an edge between the value and price instead of the edges between the signal and the value.

This contrasts sharply with the rational benchmark. There, all traders combine all available signals according to Bayes' rule, weighting each signal by its precision relative to its correlation with other signals. Information is fully utilized, and there is no role for either specialization or learning from the price.  
\subsubsection{Price and public information\label{sec:public info and price}}
In the above setting, all traders see the same signal $(s_1,s_2)$,  so it is public information. Therefore, the equilibrium price is a garbling of the two signals. A Bayesian trader recognizes this and treats price correctly as uninformative given signals she already observes. That is, conditional on $(s_1,s_2)$, any remaining variation in price reflects only the supply shock $u$.  

In contrast, CLT equilibrium need not have this feature. A fraction of traders may learn from the price instead of their private signals. A trader whose tree links both dimensions of the signal to the value neglects the (conditional) correlation between them. She double-counts their common component and overstates their precision. However, a trader whose tree  only links price to value updates correctly conditional on price. It can therefore be better to use the garbled price correctly than the raw signals incorrectly, which is what sustains learning from a redundant public price in equilibrium.

To demonstrate this formally, consider $\sigma_1=\sigma_2=\sigma_\varepsilon$. If $\bar{\rho}$ and $\sigma_u$ are small enough, then there is an equilibrium where a $\mu\in(0,1)$ fraction  condition on the tree that links both signals to $v$ (the bottom right corner of Figure \ref{fig:CLT J2}), and the remaining $(1-\mu)$ fraction condition on the tree that links everything to $p$ (the bottom left corner of Figure \ref{fig:CLT J2}). We provide the details in the appendix, and the intuition here. When all traders uses one of the two trees,   $\rho(p,v)>\rho(v,s_j)$ and $\rho(s_1,s_2)<\rho(s_j,v)$ because $\bar{\rho}$ is small. Equilibrium requires that each $s_j$ is at least as strongly correlated with $v$ as it is with $p$; otherwise, the  CLT links all variables directly to $p$ and the market cannot clear. This inequality is consistent with $\mu=1$ only when  $\sigma_u$ is large enough to hold the price's correlation with $v$ down. For small $\sigma_u$, the price is too informative about $v$ for that to be an equilibrium. We must have a positive fraction use both trees, so $\mu$ adjusts to set $\rho(p,s_j)=\rho(v,s_j)$. 

\section{Conclusion}
This paper applied a classic machine learning algorithm to a stylized model of the asset market. By construction, both leave out many features. How these simplifications compensate for one another remains an open question. In particular, CLT restricts attention to 1-order dependence trees, and the asset market has only a single endogenous random variable. A more realistic model would relax both of these. \cite{admati1985noisy}  extends the rational expectation equilibrium  to multiple assets, and the  whole price vector is typically necessary for aggregating  information about the value of any given asset.  Modern neural network models have trillions of parameters, but financial markets generate trillions of prices (when lags are taken into account). The results herein suggest that the complexity of the algorithm need not win out.
\appendix

\section{Proofs}

\subsection{Proof of Lemma \ref{lem:KLD normal}}
Let  $E=\left\{i \in \left\{ 1,\dots,n\right\}:R(i)=\emptyset \right\}$ be the ancestral nodes, and $J=\left\{ 1,\dots,n\right\} \setminus E$.
Slightly abusing notation, let $R(j)=\left\{ R(j)\right\} $ for $j\in J$.
Then we can write the density of $P_{R}$ as
\[
p_{R}(x)=\prod_{i\in E}p\left(x_{i}\right)\prod_{j\in J}p\left(x_{j}|x_{R(j)}\right)
\]
where $p$ is the density of $P$. Then, we have $D_{KL}\left(P||P_{R}\right)$ equals
\begin{align*}
& \int p(x)\ln\frac{p(x)}{p_{R}(x)}dx=-\int p(x)\ln p_{R}(x)dx-h(p)\\
 =& -\int p(x)\left(\sum_{i\in E}\ln p(x_{i})+\sum_{j\in J}\ln p\left(x_{j}|x_{R(j)}\right)\right)dx-h(p)\\
 =& -\sum_{i\in E}\int p\left(x_{i}\right)\ln p\left(x_{i}\right)dx_{i}-h(p)\\
 & \qquad-\sum_{j\in J}\int\int p\left(x_{j},x_{R(j)}\right)\left[\ln p\left(x_{j},x_{R(j)}\right)-\ln p\left(x_{R(j)}\right)\right]dx_{j}dx_{R(j)}\\
 =& \sum_{j\in J}\left[\left[\left(1+\ln2\pi\right)+\frac{1}{2}\ln\sigma_{j}^{2}\sigma_{R(j)}^{2}\left(1-\rho\left(x_{j},x_{R(j)}\right)^{2}\right)\right]
 -\left[\frac{1}{2}\left(1+\ln2\pi\right)+\frac{1}{2}\ln\sigma_{R(j)}^{2}\right]\right]\\
 & \qquad+\sum_{i\in E}\left[\frac{1}{2}\left(1+\ln2\pi\right)+\frac{1}{2}\ln\sigma_{i}^{2}\right]-h(p)\\
 =& \sum_{j\in J}\left[\frac{1}{2}\ln\left(1-\rho\left(x_{j},x_{R(j)}\right)^{2}\right)\right]+\sum_{k=1}^{n}\left[\frac{1}{2}\left(1+\ln2\pi\right)+\frac{1}{2}\ln\sigma_{k}^{2}\right]-h(p).
\end{align*}
The fourth equality comes from $p\left(x_{j}|x_{R(j)}\right)=p\left(x_{j},x_{R(j)}\right)/p\left(x_{R(j)}\right)$,
and the fifth comes from the formulae for differential entropy of
a normal distribution. Taking $\kappa=\frac{n}{2}\left(1+\ln2\pi\right)+\sum_{j=1}^{n}\frac{1}{2}\ln\sigma_{j}^{2}-h(p)$
completes the proof.

\subsection{Proof of Theorem \ref{thm: CLT exists}}

Rewrite
\[
p=\sum_{i}\sum_{j}\alpha_{j}^{i}s_{j}^{i}+\alpha_{u}u=\sum_{i}\sum_{j}\alpha_{j}^{i}\left(v+\epsilon_{j}^{i}\right)+\alpha_{u}u=\alpha_{s}\left(v+\sum_{i}\sum_{j}\hat{\alpha}_{j}^{i}\epsilon_{j}^{i}+\hat{\alpha}_{u}u\right)
\]
for $\alpha_{s}=\sum_{i}\sum_{j}\alpha_{j}^{i}$, $\hat{\alpha}_{j}^{i}=\alpha_{s}^{-1}\alpha_{j}^{i}$,
and $\hat{\alpha}_{u}=\alpha_{s}^{-1}\alpha_{u}$ when $\alpha_{s}>0$.
Observe $\sum_{i}\sum_{j}\hat{\alpha}_{j}^{i}=1$. Letting $\mathbb{I}$
be the indicator function,
\begin{align*}
E_{R}[v|s^{i},p=\alpha\cdot(s,u)] & =\frac{\sum_{j}\mathbb{I}_{R}\left(\left\{ s_{j}^{i},v\right\} \right)\sigma_{i,j}^{-2}s_{j}^{i}+\mathbb{I}_{R}\left(\left\{ p,v\right\} \right)\alpha_{s}^{-1}\sigma_{\tilde{p}}^{-2}p}{\sigma_{v}^{-2}+\sum_{j}\mathbb{I}_{R}\left(\left\{ s_{j}^{i},v\right\} \right)\sigma_{i,j}^{-2}+\mathbb{I}_{R}\left(\left\{ p,v\right\} \right)\sigma_{\tilde{p}}^{-2}}\\
Var_{R}(s^{i},p) & =\left(\sigma_{v}^{-2}+\sum_{j}\mathbb{I}_{R}\left(\left\{ s_{j}^{i},v\right\} \right)\sigma_{i,j}^{-2}+\mathbb{I}_{R}\left(\left\{ p,v\right\} \right)\sigma_{\tilde{p}}^{-2}\right)^{-1}\\
\sigma_{\tilde{p}}^{-2} & =\begin{cases}
\left(\hat{\alpha}_{-u}^{\prime}\Sigma_{\epsilon}\hat{\alpha}_{-u}+\hat{\alpha}_{u}^{2}\sigma_{u}^{2}\right)^{-1} & if\,\alpha_{s}>0\\
0 & if\,\alpha_{s}=0
\end{cases}\\
\sigma_{i,j}^{2} & =Var\left(\epsilon_{j}^{i}\right)
\end{align*}
where $\Sigma_{\epsilon}$ is the covariance matrix for $\vec{\epsilon}$.
Therefore, we have
\begin{align*}
x_{R}^{i}(s,u) & =r^{i}\sum_{j=1}^{J}\mathbb{I}_{R}\left(\left\{ s_{j}^{i},v\right\} \right)\sigma_{i,j}^{-2}s_{j}^{i}\\
 & \qquad+r^{i}\left[\mathbb{I}_{R}\left(\left\{ p,v\right\} \right)\alpha_{s}^{-1}\sigma_{\tilde{p}}^{-2}-Var_{R}(v|s^{i},p)^{-1}\right]\left(\sum_{k}\sum_{j}\alpha_{j}^{k}s_{j}^{k}+\alpha_{u}u\right)
\end{align*}
and aggregated demand equals
\[
x^{i}(s,u)=\sum_{R}\mu(R)x_{R}^{i}(s,u)=\sum_{j=1}^{J}\gamma_{j}^{i}s_{j}^{i}+\gamma_{p}^{i}\alpha_{s}^{-1}p-\beta^{i}p
\]
for
\begin{align*}
\gamma_{j}^{i} & =r^{i}\sigma_{i,j}^{-2}\sum_{R}\mu^{i}(R)\left(\sum_{j}\mathbb{I}_{R}\left(\left\{ s_{j}^{i},v\right\} \right)\right)\\
\gamma_{p}^{i} & =r^{i}\sum_{R}\mu^{i}(R)\mathbb{I}_{R}\left(\left\{ p,v\right\} \right)\sigma_{\tilde{p}}^{-2}\\
\beta^{i} & =r^{i}\sum_{R}\mu^{i}(R)\left(\sigma_{v}^{-2}+\sum_{j}\mathbb{I}_{R}\left(\left\{ s_{j}^{i},v\right\} \right)\sigma_{i,j}^{-2}+\mathbb{I}_{R}\left(\left\{ p,v\right\} \right)\sigma_{\tilde{p}}^{-2}\right)
\end{align*}
Note that $\sigma_{\tilde{p}}^{-2}$ and $\beta^{i}$ are functions
of the DM's prediction about $\alpha$, and that each $\gamma_{j}^{i}$
is a function of only $\mu^{i}$. 

We first consider $\mu$ so that $\mu^{i}(R)\geq\epsilon>0$ for all
$i$ and $R$. We show that there exists $\alpha(\vec{\mu})$ that
clears the market for any such $\vec{\mu}$. Denote $\gamma_{p}=\sum_{i}\gamma_{p}^{i}\geq0$
and $\beta=\sum_{i}\beta^{i}$, noting both depend only on $\sigma_{\tilde{p}}^{-2}$
(which depends in turn on $\alpha$) and $\vec{\mu}$. Assume $\alpha_{s}>0$
, which must be the case if $\gamma_{j}^{i}>0$ for some $i$ and
$j$ for the market to clear. This is guaranteed by $\mu^{i}(R)>0$
for all $i$ and $R$. Then, it is legitimate to write the aggregate
demand as
\[
x(s,u)=\sum_{i}\sum_{j}\left(\gamma_{j}^{i}+\gamma_{p}\alpha_{s}^{-1}\alpha_{j}^{i}-\beta\alpha_{j}^{i}\right)s_{j}^{i}+\left[\gamma_{p}\alpha_{s}^{-1}-\beta\right]\alpha_{u}u.
\]
Since $x(s,u)=u$, we must have
\begin{align*}
\gamma_{j}^{i}+\left(\gamma_{p}\alpha_{s}^{-1}-\beta\right)\alpha_{j}^{i} & =0\\
\frac{\gamma_{j}^{i}}{\beta-\gamma_{p}\alpha_{s}^{-1}} & =\alpha_{j}^{i}\\
\left[\gamma_{p}\alpha_{s}^{-1}-\beta\right]\alpha_{u} & =1
\end{align*}
This implies that 
\[
-\frac{\alpha_{j}^{i}}{\alpha_{u}}=\gamma_{j}^{i}=\sigma_{i,j}^{-2}r^{i}\mu^{i}\left(\left\{ R:\left\{ s_{j}^{i},v\right\} \in R\right\} \right).
\]
Adding up,
\begin{equation}
-\frac{\alpha_{u}}{\alpha_{s}}=\left(\sum_{i,j}r^{i}\mu^{i}\left(\left\{ R:\left\{ s_{j}^{i},v\right\} \in R\right\} \right)\sigma_{i,j}^{-2}\right)^{-1}\equiv h\label{eq:Eqbn h}
\end{equation}
which yields
\begin{equation}
\hat{\alpha}_{j}^{i}=\frac{\alpha_{j}^{i}}{\alpha_{s}}=\frac{\gamma_{j}^{i}}{\sum_{i^{\prime},j^{\prime}}\gamma_{j^{\prime}}^{i^{\prime}}}=\frac{\sigma_{i,j}^{-2}r^{i}\mu^{i}\left(\left\{ R:\left\{ s_{j}^{i},v\right\} \in R\right\} \right)}{\sum_{i^{\prime},j^{\prime}}r^{i^{\prime}}\mu^{i^{\prime}}\left(\left\{ R:\left\{ s_{j}^{i^{\prime}},v\right\} \in R\right\} \right)\sigma_{i^{\prime},j^{\prime}}^{-2}}.\label{eq:alpha hat}
\end{equation}
Writing out $\gamma=\left(\frac{\gamma_{j}^{i}}{\sum_{i^{\prime},j^{\prime}}\gamma_{j^{\prime}}^{i^{\prime}}}\right)_{i,j}$
as an appropriately ordered vector, 
\begin{equation}
\sigma_{\tilde{p}}^{-2}=\left(\gamma^{\prime}\Sigma_{\epsilon}\gamma+h^{2}\sigma_{u}^{2}\right)^{-1};\label{eq:Eqbn sigma p}
\end{equation}
that is, $\sigma_{\tilde{p}}^{-2}$ is determined uniquely by $h$
and $\gamma$, so it is uniquely determined by $\vec{\mu}$. Since
$\beta^{i}$ and $\gamma_{p}^{i}$ depend only on $\mu$ and $\sigma_{\tilde{p}}^{-2}$,
they are also uniquely determined by $\vec{\mu}$. Finally, notice
that 
\begin{align*}
1 & =\alpha_{u}\left[\gamma_{p}\alpha_{s}^{-1}-\beta\right]=-h\gamma_{p}-\beta\alpha_{u}
\end{align*}
pins down $\alpha_{u}$ uniquely, determining a unique $\alpha\left(\vec{\mu}\right)$
that solves the above system.

On the other hand, no equilibrium exists if $\gamma_{j}^{i}=0$ for
all $i$ and $j$. For then, 
\[
x(s,u)=\sum\left(\gamma_{p}\alpha_{s}^{-1}\alpha_{j}^{i}-\beta\alpha_{j}^{i}\right)s_{j}^{i}+\left[\gamma_{p}\alpha_{s}^{-1}-\beta\right]\alpha_{u}u.
\]
Equilibrium requires $\gamma_{p}\alpha_{s}^{-1}=\beta$ so that aggregate
demand does not depend on $s_{j}^{i}$, and also that $\gamma_{p}\alpha_{s}^{-1}\neq\beta$
so that aggregate demand equals $u$. Clearly, this is a contradiction.
This cannot happen when $\mu^{i}(R)\geq\epsilon>0$ for all $R$,
since then $\gamma_{j}^{i}\geq r^{i}\epsilon\sigma_{i,j}^{-2}>0$.

Fix $\epsilon\in\left(0,\min_{i}\frac{1}{|\mathcal{R}^{i}|}\right)$,
and 
\[
\mathcal{M}_{\epsilon}=\left\{ \left(\mu^{1},\dots,\mu^{I}\right)\in\prod\Delta\mathcal{R}^{i}:\mu^{i}(R)\geq\epsilon\forall R\in\mathcal{R}^{i}\forall i\right\} .
\]
Note that $\gamma_{j}^{i}>0$ whenever $\vec{\mu}\in\mathcal{M}_{\epsilon}$,
so a unique $\alpha\left(\vec{\mu}\right)$ exists that clears the
market. Moreover, $\alpha$ is continuous by the implicit function
theorem. We search for an $\epsilon$ equilibrium, 
\[
\mu^{i}(R)>\epsilon\implies R\in\arg\min_{R'\in\mathcal{R}^{i}}
D_{KL}\left(\nu^{i}(\cdot|\alpha(\vec{\mu}))||\nu_{R}^{i}(\cdot|\alpha(\vec{\mu}))\right)
\]
for all $i$. Consider the correspondence $\Gamma$ where 
\[
\Gamma(\mu)=
\left\{ \vec{m}\in\mathcal{M}_{\epsilon}:
m^{i}(R)>\epsilon\implies 
R\in\arg\min_{R'\in\mathcal{R}^{i}}
D_{KL}\left(\nu^{i}(\cdot|\alpha(\vec{\mu}))||\nu_{R}^{i}(\cdot|\alpha(\vec{\mu}))\right)\right\} .
\]
$\Gamma$ has a closed graph and is non-empty and convex-valued. Since
$\mathcal{M}_{\epsilon}$ is compact and convex, and each $\mathcal{R}^{i}$
is finite, Kakutani implies that a fixed point exists; call it $\mu_{\epsilon}$.

Consider a sequence $\epsilon_{1},\epsilon_{2},\dots$ with $\epsilon_{t}\rightarrow0$
so that $\mu_{t}=\mu_{\epsilon_{t}}$ converges to $\mu^{*}$, respectively.
This subsequence exists because $\mathcal{M}_{0}\supset\mathcal{M}_{\epsilon}$
is compact. We show that the pair $\left(\mu^{*},\alpha^{*}=\alpha(\mu^{*})\right)$
corresponds to an equilibrium.

Let $\alpha_{t}=\alpha(\mu_{t})$. Define $\mathcal{R}_{p}^{i}=\left\{ R\in\mathcal{R}^{i}:\{z,v\}\in R\implies z=p\right\} $,
the set of trees that only use the price to learn about $v$. For
contradiction, suppose that $\mu^{*}(\mathcal{R}_{p}^{i})=1$ for
all $i$. Then, $\gamma_{j}^{i}\left(\mu_{t}\right)\rightarrow0$
for all $i$ and $j$ which in turn implies $\frac{\left(\alpha_{t}\right)_{j}^{i}}{\left(\alpha_{t}\right)_{u}}\rightarrow0$.
But then $\rho(p,v),\rho(p,s_{j}^{i})\rightarrow0$ for all $i,j$
and in particular for $t$ large, $\rho(p,v),\rho\left(s_{j}^{i},p\right)<\rho\left(s_{j}^{i},v\right)$
for all $i,j$, and so $\mu_{t}^{i}(R^{\prime})<\epsilon_{t}$ for
all $R^{\prime}\in\mathcal{R}_{p}^{i}$, a contradiction. This means
that $\mu^{i*}(\mathcal{R}_{p}^{i})<1$ for some $i$, and therefore
$\gamma_{j}^{i}\left(\mu^{*}\right)>0$ for at least one $i,j$ pair. Consequently, $\alpha\left(\mu^{*}\right)$ is well-defined and $\mu^* \in \mathcal{M}_\varepsilon$ for some $\varepsilon>0$. Consequently, 
$\alpha_{t}\rightarrow\alpha\left(\mu^{*}\right)=\alpha^{*}$. By
the above, $\alpha^{*}$ clears the market for $\mu^{*}$, and by
the Berge maximum theorem, $\mu^{*}$ attaches probability 1 to divergence
minimizing trees.

\subsection{Proof of Theorem \ref{thm: info efficiency}}

Fix $I$, $\sigma_{v}$, and $\Sigma^{i}$. Note that
\[
\left(\begin{array}{c}
v\\
\vec{s}
\end{array}\right)\sim N\left(0,\left[\begin{array}{cc}
\sigma_{v}^{2} & \Sigma_{S,v}\\
\Sigma_{v,S} & \Sigma_{S}
\end{array}\right]\right)
\]
 where $\Sigma_{S}$ is a $IJ\times IJ$ matrix, $\Sigma_{v,S}=\Sigma_{S,v}^{\prime}$
is $IJ\times1$ matrix. Then 
\[
v|\vec{s}\sim N(\Sigma_{S,v}\Sigma_{S}^{-1}\vec{s},\sigma_{v}^{2}-\Sigma_{S,v}\Sigma_{S}^{-1}\Sigma_{v,S})
\]
 by the projection theorem, and $v\perp\vec{s}|E[v|\vec{s}]$. Denote by $\mathcal{I}(\cdot)$ the mutual information between two variables. Letting
$f^{*}(\vec{s})=\Sigma_{S,v}\Sigma_{S}^{-1}\vec{s}=E[v|\vec{s}]$,
for any $f:\mathbb{R}^{(IJ)^{2}}\rightarrow\mathbb{R}$, $v\perp f(\vec{s})|f^{*}(\vec{s})$,
so
\[
I^{*}\equiv \mathcal{I}(v,f^{*})\geq \mathcal{I}(v,f)
\]
by Theorem 2.8.1 of Cover and Thomas, with equality only if $v\perp f^*(\vec{s})|f(\vec{s})$ which requires that $f=h\circ f^*$ for some bijection $h:\mathbb{R}\rightarrow \mathbb{R}$.

Let $\Delta=\left[-1,1\right]^{IJ+1}$. Any $\alpha$ can be rescaled
to $\alpha^{\prime}\in\Delta$ without changing the mutual information
between any pair of variables. Let $p\left(\alpha,\sigma_{u}\right)$ be a random variable equal to $\alpha\cdot(\vec{s},u)$
in distribution, given that $u\sim N\left(0,\sigma_{u}^{2}\right)$
independently of the other variables.  Let
\[
g=\left(\alpha,\sigma_{u}\right)\mapsto \mathcal{I}\left(v,p\left(\alpha,\sigma_{u}\right)\right).
\]
and  $C \equiv g^{-1}(I^{*})$. Note $C$ is closed since $g$ is the composition of
continuous functions and non-empty since $\left(\Sigma_{S,v}\Sigma_{S}^{-1},0;\sigma_{u}\right)\in C$
for every $\sigma_{u}$. Note $\mathcal{I}(v,p(\cdot))$, $\mathcal{I}(p(\cdot),s_{j}^{i})$,
and $\mathcal{I}(v,s_{j}^{i})$ are all continuous in $\left(\alpha,\sigma_{u}\right)$.
Moreover, for any $\left(\alpha,\sigma_{u}\right)\in C$ and $(i,j)$ pair, $\mathcal{I}\left(s_{j}^{i},p\left(\alpha,\sigma_{u}\right)\right)>\mathcal{I}(s_{j}^{i},v)$
and $\mathcal{I}\left(v,p\left(\alpha,\sigma_{u}\right)\right)>\mathcal{I}\left(v,s_{j}^{i}\right)$.
This follows from $\mathcal{I}\left(v,p\left(\alpha,\sigma_{u}\right)\right)=\mathcal{I}(v,f^{*})$
if and only if $p\left(\alpha,\sigma_{u}\right)$ is a sufficient
statistic for $v$ given $\vec{s}$, and for any $\left(\alpha^{*},\sigma_{u}^{*}\right)\in C$,
$\mathcal{I}\left(v,p\left(\alpha,\sigma_{u}\right)\right)> \mathcal{I}(v,s_{j}^{i})$
and $\mathcal{I}\left(s_{j}^{i},p\left(\alpha^*,\sigma^*_{u}\right)\right)> \mathcal{I}(s_{j}^{i},v)$
by Theorem 2.8.1 of Cover and Thomas. Set 
\[
G=\left(\alpha,\sigma_{u}\right)\mapsto\left(\mathcal{I}\left(v,p\left(\alpha,\sigma_{u}\right)\right)-\mathcal{I}(v,s_{j}^{i}),\mathcal{I}\left(s_{j}^{i},p\left(\alpha,\sigma_{u}\right)\right)-\mathcal{I}(s_{j}^{i},v)\right)_{i,j}
\]
Since $C$ is a closed subset of the compact set $\Delta\times[0,1]$, $G$ is continuous,   and $G\left(\alpha,\sigma_{u}\right) \gg 0$
for all $\left(\alpha,\sigma_{u}\right)\in C$, there exists $c>0$
so that $G\left(\alpha,\sigma_{u}\right)\geq\left(c,c,\dots,c\right)$
for all $\left(\alpha,\sigma_{u}\right)\in C$ by the Weierstraus
theorem.

Since $\Delta\times[0,1]$ is compact, we can take $G$ to be uniformly
continuous, so there exists $\epsilon>0$ so that $G\left(a,\sigma_{u}\right)>\left(\frac{c}{2},\frac{c}{2},\dots,\frac{c}{2}\right)$
for every $\left(a,\sigma_{u}\right)\in\cup_{\left(a^{\prime},\sigma_{u}\right)\in C}B_{\epsilon}\left(a^{\prime},\sigma_{u}\right)=O$.
  We claim that there exists $\delta^{*}>0$ so that $g\left(a,\sigma_{u}\right)>I^{*}-\delta^{*}$
implies that $\left(a,\sigma_{u}\right)\in O$. If not, then for all
$n\in\mathbb{N}$, $\exists y_{n}\in\Delta\times[0,1]$ so that $g(y_{n})>I^{*}-\frac{1}{n}$
and $y_{n}\notin O$ for all $n$. For some $y^{*}\in\Delta\times[0,1]$,
$y_{n}\rightarrow y^{*}$ (after taking a subsequence). By continuity
and that $g(y_{n})\in [I^*-\tfrac{1}{n}, I^{*}]$, $g(y^{*})=I^{*}$. But then $y^{*}\in C\subset O$,
so $y_{n}\in O$ for all $n$ large enough, a contradiction.

Now, if $\left(\alpha,\sigma_{u}\right)$ is such that $g\left(\alpha,\sigma_{u}\right)>I^{*}-\delta^{*}$,
the CLT $R_{i}$ for $i$ is such that $\{v,x\}\in R_{i}$ if and
only if $x=p$. The edge with highest weight involving either $v$
or $p$ is $\left\{ v,p\right\} $, so $\left\{ v,p\right\} \in R_{i}$.
Replacing the edge $\left\{ s_{j}^{i},v\right\} $ with the edge $\left\{ s_{j}^{i},p\right\} $
decreases the divergence of the tree. This creates a cycle only if
there are $k_{1},\dots,k_{m}$ so that $\left\{ s_{j}^{i},s_{k_{1}}^{i}\right\} ,\dots,\left\{ s_{k_{m-1}}^{i},s_{k_{m}}^{i}\right\} ,\left\{ s_{k_{m}}^{i},v\right\} \in R_{i}$
or $\left\{ s_{j}^{i},s_{k_{1}}^{i}\right\} ,\dots,\left\{ s_{k_{m-1}}^{i},s_{k_{m}}^{i}\right\} ,\left\{ s_{k_{m}}^{i},p\right\} \in R_{i}$.
Since $\left\{ v,p\right\} \in R_{i}$, $\left\{ s_{j}^{i},v\right\} $
would also create a cycle, so the replacement does not create a cycle.
Therefore, $\left\{ v,p\right\} $ is the only edge $R_{i}$ containing
$v$. However, this implies that $E_{R_{i}}[v|s^{i},p]=E[v|p]$ for
all $i$, which means that demand does not depend on $s^{i}$, making
market clearing impossible with $\sigma_{u}>0$.

Using formulas for normal distribution, $I^{*}=-(1-\rho(v,f^{*})^{2})$
and $I(v,f)=-(1-\rho(v,f)^{2})$, so $\rho(v,f)^{2}\leq\rho(v,f^{*})^{2}$.
In particular, since $p$ is a function of $(\vec{s},u)$ and $u$
is independent of $\left(v,\vec{s}\right),$ we have $v\perp p|f(\vec{s})$,
so $\rho(v,p)^{2}\leq\rho(v,f^{*})^{2}$, and there exists $\delta>0$
so that if $\rho(v,p)^{2}>\rho(v,f^{*})^{2}-\delta$, then $I(v,p)>I^{*}-\delta^{*}$,
and so $p$ cannot be an equilibrium.

\subsection{Proof of Theorem \ref{thm: CLT J=1}}
By Theorem \ref{thm: CLT exists},
\[
x(s,u)=\sum_{i}\left(\gamma^{i}+\gamma_{p}\alpha_{s}^{-1}\alpha^{i}-\beta\alpha^{i}\right)s^{i}+\left[I\gamma_{p}\alpha_{s}^{-1}-\sum\beta^{i}\right]\alpha_{u}u
\]
 where 
\begin{align*}
\gamma^{i} & =r\sigma_{\varepsilon}^{-2}\left(1-\mu^{i}(R_{p})\right)\\
\gamma_{p} & =\sum_{i}r\sigma_{\tilde{p}}^{-2}\left(1-\mu^{i}(R_{s})\right)\\
\beta & =\sum_{i}r\left(\sigma_{v}^{-2}+\sigma_{\tilde{p}}^{-2}\left(1-\mu^{i}(R_{s})\right)+\sigma_{\varepsilon}^{-2}\left(1-\mu^{i}(R_{p})\right)\right).
\end{align*}
Suppose that $\mu^{i}=\mu$ for all $i$. Since $x(s,u)=u$, substituting
into Equations (\ref{eq:Eqbn h}) and (\ref{eq:Eqbn sigma p}) gives
\begin{align*}
h & =-\frac{\alpha_{u}}{\alpha_{s}}=\frac{\sigma_{\varepsilon}^{2}}{I r\left(1-\mu(R_{p})\right)}\\
\sigma_{\tilde{p}}^{-2} & =\left(\gamma^{\prime}\Sigma_{\epsilon}\gamma+h^{2}\sigma_{u}^{2}\right)^{-1}=\left(I^{-1}\sigma_{\varepsilon}^{2}+\frac{\sigma_{\epsilon}^{4}}{r^{2}I^2 \left(1-\mu(R_{p})\right)^{2}}\sigma_{u}^{2}\right)^{-1}.
\end{align*}
Notice that, independent of $\mu$ and $\alpha$, we have
\[
\rho(v,p)\leq\rho(p,s^{i})\iff\left(I^{2}-2I\right)\sigma_{v}^{2}\leq\sigma_{\varepsilon}^{2}.
\]
 If $\rho(v,p)<\rho(p,s^{i})$, then $\mu(R_{v})=0$, and equilibrium
requires that $\rho(v,s^{i})\geq\rho(v,p)$ since otherwise $\mu(R_{p})=1$.
Now, $\rho(v,p)\leq\rho(v,s^{i})$ holds if and only if 
$1-\mu(R_{p})\leq\sqrt{\frac{\sigma_{u}^{2}\sigma_{\varepsilon}^{2}}{I\left(I-1\right)r^{2}}}$.
So if $\sqrt{\frac{\sigma_{u}^{2}\sigma_{\varepsilon}^{2}}{I\left(I-1\right)r^{2}}}\geq 1$,
$\mu(R_{s})=1$ is an equilibrium. Otherwise, $1-\mu(R_{p})=\sqrt{\frac{\sigma_{u}^{2}\sigma_{\varepsilon}^{2}}{I\left(I-1\right)r^{2}}}=\mu(R_{s})$
characterizes the equilibrium. 

Moreover, when $0<\mu(R_{p})$,
$Var_{R_{s}}(v|s,p)=Var_{R_{p}}(v|s,p)=\frac{\sigma_{v}^{2}\sigma_{\varepsilon}^{2}}{\sigma_{v}^{2}+\sigma_{\varepsilon}^{2}}=rI\left(\beta\right)^{-1}$.
Denote the equilibrium price $p^{s}=\alpha_{s}^{s} I^{-1}\sum s^{i}+\alpha_{u}^{s}u$.
From above we have
\[h  =\sqrt{\frac{I-1}{I}}\frac{\sigma_{\varepsilon}}{\sigma_u}\]
Solving  for $\alpha_{s}^{s}$
gives 
\[
\alpha_{s}^{s}=\frac{\sigma_{\varepsilon}^{-2}}{\sigma_{v}^{-2}+\sigma_{\varepsilon}^{-2}}
\]
Using the above formulas for $h$ and $\mu(R_{p})$, we have
\begin{align*}
h^{2}\sigma_{u}^{2} & =\frac{I-1}{I} \sigma_{\varepsilon}^2
\end{align*}
and therefore 
\begin{align*}
Var(p^{s}) & =\left(\alpha_{s}^{s}I^{-1}\right)^{2}\left(I^{2}\sigma_{v}^{2}+I\sigma_{\varepsilon}^{2}\right)+\left(\alpha_{s}^{s}h\right)^{2}\sigma_{u}^{2}.\\
 & =\frac{\sigma_{\varepsilon}^{-4}}{\left(\sigma_{v}^{-2}+\sigma_{\varepsilon}^{-2}\right)^{2}}\left(\sigma_{v}^{2}+I^{-1}\sigma_{\varepsilon}^{2}+\frac{I-1}{I} \sigma_{\varepsilon}^2\right)
\end{align*}

If $\rho(v,p)>\rho(p,s^{i})$, then $\mu(R_{s})=0$. Equilibrium requires
that $\rho(v,s^{i})\geq\rho(p,s^{i})$ since otherwise $\mu(R_{p})=1$.
Now, $\rho(s^{i},p)\leq\rho(v,s^{i})$ holds if and only if $\frac{\sigma_{\varepsilon}^{2}\sigma_{u}^{2}\sigma_{v}^{2}}{r^{2}\left(\sigma_{\varepsilon}^{2}+I\sigma_{v}^{2}\right)}\geq\left(1-\mu(R_{p})\right)^{2}$.
So if $\frac{\sigma_{\varepsilon}^{2}\sigma_{u}^{2}\sigma_{v}^{2}}{r^{2}\left(\sigma_{\varepsilon}^{2}+I\sigma_{v}^{2}\right)}\geq1$,
$\mu(R_{v})=1$ is an equilibrium. Otherwise, $\sqrt{\frac{\sigma_{\varepsilon}^{2}\sigma_{u}^{2}\sigma_{v}^{2}}{r^{2}\left(\sigma_{\varepsilon}^{2}+I\sigma_{v}^{2}\right)}}=\mu(R_{v})=1-\mu(R_{p})$
characterizes the equilibrium.

Moreover, when $\mu(R_p)>0$,
\[
\tau_{R_{v}}^{-1}=Var_{R_{v}}(v|s,p)=\left(\sigma_{v}^{-2}+\sigma_{\varepsilon}^{-2}+\sigma_{\tilde{p}}^{-2}\right)^{-1}<\left(\sigma_{v}^{-2}+\sigma_{\tilde{p}}^{-2}\right)^{-1}=Var_{R_{p}}(v|s,p)=\tau_{R_{p}}^{-1}
\]
Denote the equilibrium price $p^{v}=\alpha_{s}^{v}\left(I^{-1}\sum s^{i}\right)+\alpha_{u}^{v}u$.
Then, 
\[
x^{i}(s^{i},p)=r(1-\mu(R^{p}))\sigma^{-2}_{\varepsilon}s^{i}+r\left(\sigma_{\tilde{p}}^{-2}\left(\alpha_{s}^{v}\right)^{-1}-\tau\right)p
\]
with $$\tau=\mu(R_{p})\tau_{R_{p}}v+(1-\mu(R_{p}))\tau_{R_{v}} =\sigma_{\tilde{p}}^{-2}+\sigma_{v}^{-2}+(1-\mu(R_{p}))\sigma_{\varepsilon}^{-2}.$$
As $\sigma_{u}^{2}\rightarrow0$, $\mu(R_{p})\rightarrow1$ so
\begin{align*}
\alpha_{s}^{v} & =\frac{(1-\mu(R_{p}))\sigma_{\varepsilon}^{-2}+\sigma_{\tilde{p}}^{-2}}{\sigma_{\tilde{p}}^{-2}+\sigma_{v}^{-2}+(1-\mu(R_{p}))\sigma_{\varepsilon}^{-2}}\\
 & \rightarrow\frac{\sigma_{\tilde{p}}^{-2}}{\sigma_{v}^{-2}+\sigma_{\tilde{p}}^{-2}}
\end{align*}
Also,
\[
h(\sigma_{u}^{2})^{2}\sigma_{u}^{2}=\left(\frac{\sigma_{\varepsilon}^{2}}{r\left(1-\mu(R_{p})\right)}\right)^{2}\sigma_{u}^{2}=\frac{r^{2}\left(I^{-1}\sigma_{\varepsilon}^{2}+\sigma_{v}^{2}\right)}{I\sigma_{\varepsilon}^{2}\sigma_{u}^{2}\sigma_{v}^{2}}\left(\frac{\sigma_{\varepsilon}^{2}}{r}\right)^{2}\sigma_{u}^{2}=\sigma_{\varepsilon}^{2}\frac{I^{-1}\sigma_{\varepsilon}^{2}+\sigma_{v}^{2}}{I\sigma_{v}^{2}}
\]
so 
\begin{align*}
Var(p^{v}) & \rightarrow\left(\frac{\sigma_{\tilde{p}}^{-2}}{\sigma_{\tilde{p}}^{-2}+\sigma_{v}^{-2}}\right)^{2}\left(\sigma_{v}^{2}+I^{-1}\sigma_{\varepsilon}^{2}+\sigma_{\varepsilon}^{2}\frac{I^{-1}\sigma_{\varepsilon}^{2}+\sigma_{v}^{2}}{I\sigma_{v}^{2}}\right).
\end{align*}
Since $\rho(v,p)>\rho(s^{i},p)=\rho(s^{i},v)$, $\sigma_{\tilde{p}}^{-2}>\sigma_{\varepsilon}^{-2}$,
and so 
\[
\lim_{\sigma_{u}^{2}\rightarrow0}Var(p^{v})\geq\left(\frac{\sigma_{\varepsilon}^{-2}}{\sigma_{\varepsilon}^{-2}+\sigma_{v}^{-2}}\right)^{2}\left(\sigma_{v}^{2}+I^{-1}\sigma_{\varepsilon}^{2}+\frac{I\sigma_{\varepsilon}^{-2}+\sigma_{v}^{-2}}{I^{2}\sigma_{\varepsilon}^{-2}}\right).
\]

\subsection{Proof of Corollary \ref{cor: rate of convergence}}
Let $K=\frac{r^{2}}{\sigma_{u}^{2}\sigma_{\varepsilon}^{2}}$ and $V(I)=K^{-1/3}I^{-1/3}$. Claim that $(1-\varphi) / V(I) \rightarrow 1$. This follows from
\[
\frac{1-V(I)}{V(I)^3} \frac{(1-\varphi)^3}{\varphi}=\frac{1-K^{-1/3}I^{-1/3}}{K I^{-1}}\frac{1}{(I-1)K}=\frac{I-K^{-1/3}I^{2/3}}{(I-1)}
\rightarrow 1\]
For $I$, the extra variance in $p^h/{\alpha^h_s} $ conditional on $v$ relative to $E[v|\vec{s}]$ is
\[
\phi^h(I)=\left(\frac{\alpha_u^{h}}{\alpha_s^h} \right)^{2}\sigma_{u}^{2}=\frac{\sigma_{\varepsilon}^{4}\sigma_{u}^{2}}{I^{2}r^{2}(1-\varphi)^{2}}
\]
By the above, $\phi^h(I)=O(I^{-4/3})$. Since $$\rho\left(\mathbb{E}[v|(s_i)_{i=1}^I],v \right)^2=\frac{\sigma_v^2}{\sigma_v^2+I^{-1}\sigma_\varepsilon^2}=\frac{\sigma_v^2 + \frac{\sigma_v^2}{\sigma_v^2+I^{-1}\sigma_\varepsilon^2}A}{\sigma_v^2+I^{-1}\sigma_\varepsilon^2+A}$$ for any $A\geq 0$, we have 
\begin{align*}
\rho\left(\mathbb{E}[v|(s_i)_{i=1}^I],v \right)^2-\rho(p^h,v)^2=&
\frac{\sigma_v^2+\phi^h(I)\frac{\sigma_v^2}{\sigma_v^2+I^{-1}\sigma_\varepsilon^2}}{\sigma_v^2+I^{-1}\sigma_\varepsilon^2 +\phi^h(I)}
-\frac{ \sigma_v^2}{\sigma_v^2+I^{-1}\sigma_\varepsilon^2+\phi^h(I)}\\
=&
\frac{  \phi^h(I)\frac{2\sigma_v^2+I^{-1}\sigma_\varepsilon^2}{\sigma_v^2+I^{-1}\sigma_\varepsilon^2}}{\sigma_v^2+I^{-1}\sigma_\varepsilon^2+\phi^h(I)}
=O(I^{-4/3}).
\end{align*}

By Theorem \ref{thm: CLT J=1}, for $I$ large, $\rho(v,p^{CLT})=\frac{\sigma_v^2}{\sigma_v^2+I^{-1}\sigma_\varepsilon^2}=\rho\left(\mathbb{E}[v|(s_i)_{i=1}^I],v \right)^2$. Therefore, \[
\rho\left(\mathbb{E}[v|(s_i)_{i=1}^I],v \right)^2-\rho(v,p^{CLT})^2=\frac{\left(\sigma_v^2+I^{-1}\sigma_\varepsilon^2\right)\sigma_v^2-\sigma_v^4}{\left(\sigma_v^2+I^{-1}\sigma_\varepsilon^2\right)^2}=\frac{I^{-1}\sigma_\varepsilon^2}{\left(\sigma_v^2+I^{-1}\sigma_\varepsilon^2\right)^2}=O(I^{-1}).
\]
Conclude $\rho(v,p^{CLT})$ converges to $\rho^*$ slower than does $\rho(v,p^{h})$.
\subsection{Proof of Theorem \ref{thm: specialization}}

The correlations are
\begin{align*}
\rho(s_{1}^{i},s_{2}^{i}) & =\frac{\sigma_{v}^{2}+\bar{\rho}\sigma_{1}\sigma_{2}}{\sqrt{(\sigma_{v}^{2}+\sigma_{1}^2 )(\sigma_{v}^{2}+\sigma^2_{2})}}\\
\rho(s_{j}^{i},v) & =\frac{\sigma_{v}}{\sqrt{\sigma_{v}^{2}+\sigma_{j}^{2}}}
\end{align*}
Note $\rho(s^i_1,v) \geq \rho(s^i_2,v)$ and  
\[
\rho(s^i_{1},s^i_{2}) \geq  \rho(s^i_{2},v)\iff \bar{\rho} \geq \hat{\rho}\equiv \frac{\sqrt{(\sigma_{v}^{2}+\sigma_{1}^2 )}\sigma_v-\sigma_{v}^{2}}{\sigma_{1}\sigma_{2}}.
\]
For a subset $I\subset \{s_1,s_2,p\}$, denote $$R_I=\{R \in \mathcal{R}: [(v,i)\in R \ or\ (i,v) \in R]\iff i \in I\}.$$
Suppose that $\bar{\rho} > \hat{\rho}$, so $0<\rho(s_2,v)<\rho(s_1,s_2)$. Consider a DAG $R \in R_{\{s_1,s_2\}}\cup R_{\{s_1,s_2,p\}}$. Let $R'$ remove the edge $\{v,s_2\}$ and add the edge $\{s_1,s_2\}$. This creates a cycle only if $\{s_2,p\}$ and $\{s_1,p\} $ belong to $ R$. But then $v-s_1-p-s_2-v$ is a cycle in $R$. Moreover, by Lemma \ref{lem:KLD normal}, $D_{KL}(\nu,\nu_{R'})<D_{KL}(\nu,\nu_{R})$.  Therefore, $\mu(R)=0$.

\subsection{Details for Section \ref{sec:public info and price}}
Adopt the notation convention in the proof of Theorem \ref{thm: specialization}. Assume an equilibrium price of the form $p=\alpha_s \frac12 \left( s_1 +s_2\right)+\alpha_u u$. Then, 
\begin{align*}
\rho(s_{1},s_{2}) & <\rho(s_{j},v)\iff\frac{\sigma^{2}_{v}+\bar{\rho}\sigma^{2}_{\varepsilon}}{\sqrt{\sigma^{2}_{v}+\sigma^{2}_{\varepsilon}}}<\sigma_{v}
\end{align*}
and
\[
\rho(s_{j},p)<\rho(v,p)\iff\frac{\sigma^{2}_{v}+(\frac12+\bar{\rho}\frac12)\sigma^{2}_{\varepsilon}}{\sqrt{\sigma^{2}_{v}+\sigma^{2}_{\varepsilon}}}<\sigma_{v}.
\]
Since $\frac12+\bar{\rho}\frac12 \geq \bar{\rho}$, both inequalities hold when 
\[
\bar{\rho} < 2\frac{\sigma_{v}\sqrt{\sigma^{2}_{v}+\sigma^{2}_{\varepsilon}}-\sigma^{2}_{v}}{\sigma^{2}_{\varepsilon}} -1,
\]
i.e., when $\bar{\rho}$ sufficiently close to $-1$. These are independent of $\alpha_s$ and $\alpha_u$.  In what follows, we consider $\bar{\rho}$ satisfying the above inequality.

If 
\(
\rho(s_{j},p)>\rho(v,s_{j})
\),
then $\mu(R_{\{p\}})=1$ since the divergence minimizing tree must belong to $R_p$. But everyone using $R\in R_{\{p\}}$ cannot be an equilibrium. 
We must instead have that
\[
\rho(v,s_{j})\geq\rho(p,s_{j})
\]
By applying the greedy algorithm, we see that the divergence minimizing trees include those in $R_{\{s_1,s_2\}}$ and, when the inequality is not strict, those is $R_{\{p\}}$. 
Suppose that $\mu(R_{\{s_1,s_2\}})=\mu=1-\mu(R_{\{p\}})$.
By Equations (\ref{eq:Eqbn h}) and (\ref{eq:Eqbn sigma p})
\[
h=(2r\mu)^{-1}\sigma^{2}_{\varepsilon},
\]
so letting $q=\frac12+\frac12\bar{\rho}$,
\[
\rho(v,s_j)=\frac{\sigma_{v}}{\sqrt{\sigma^{2}_{v}+\sigma^{2}_{\varepsilon}}}
\geq
\frac{\sigma_{v}^2+q\sigma_\varepsilon^2}{\sqrt{\sigma^{2}_{v}+\sigma^{2}_{\varepsilon}}\sqrt{\sigma^{2}_{v}+q\sigma^{2}_{\varepsilon}+(2r\mu)^{-2}\sigma^{4}_{\varepsilon}\sigma^{2}_{u}}}=\rho(p,s_j).
\]
Therefore, equilibrium is characterized by\[
\sigma_{v}
\geq
\frac{\sigma_{v}^2+q\sigma_\varepsilon^2}{\sqrt{\sigma^{2}_{v}+q\sigma^{2}_{\varepsilon}+(2r\mu)^{-2}\sigma^{4}_{\varepsilon}\sigma^{2}_{u}}}\]
with equality whenever $\mu<1$. Note that equality must obtain when $\sigma_u$ is sufficiently small. The right-hand side increases in $\sigma_u$ and decreases in $\mu$. When  $\mu=1$, it exceeds $\sigma_v$ for $\sigma_u=0$ and so also for small $\sigma_u$. We can find $\mu\in (0,1)$ that sets them equal since it approaches $0$ as $\mu\rightarrow 0$.

%

\bibliographystyle{plainnat}
\bibliography{bibliography}

\end{document}